\documentclass[11pt,a4paper]{article}

\usepackage{graphicx}
\usepackage{afterpage}
\usepackage{epsfig,cite}
\usepackage{amssymb}
\usepackage{amsmath}
\usepackage{dsfont}
\usepackage{multirow}
\usepackage{url,hyperref}
\usepackage{bm}

\usepackage{url}
\usepackage{hyperref}

\usepackage{color}
\definecolor{comment}{rgb}{0,0.3,0}
\definecolor{identifier}{rgb}{0.0,0,0.3}

\usepackage{listings}

\definecolor{listinggray}{gray}{0.9}
\definecolor{lbcolor}{rgb}{0.9,0.9,0.9}
\begin{document}

%%%%%%%%%%%%%%%%%%%%%%%%%%%%%%%%%%%%%%%%%%%%%%%%
\begin{figure}[h]
\epsfig{width=0.30\textwidth,figure=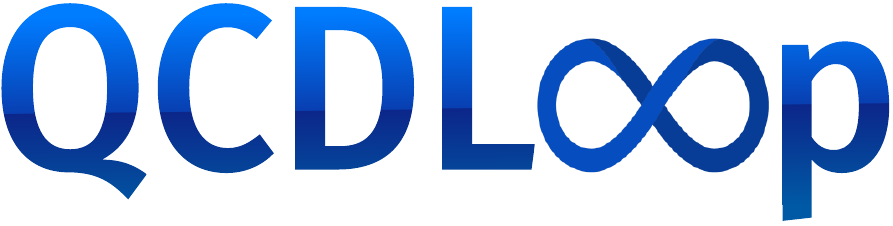}
\end{figure}
%%%%%%%%%%%%%%%%%%%%%%%%%%%%%%%%%%%%%%%%%%%%%%%%%

\vspace{-2.0cm}
\begin{flushright}
  CERN-TH-2016-101\\
  IPPP/16/29\\
\end{flushright}
\vspace{1cm}

\begin{center}
{\Large \bf {\tt QCDLoop}: a comprehensive framework for one-loop scalar integrals}
\vspace{.7cm}

Stefano~Carrazza$^{1}$, R.~Keith~Ellis$^{2}$ and Giulia~Zanderighi$^{1,3}$

\vspace{.3cm}
{\it ~$^1$ Theoretical Physics Department, CERN, Geneva, Switzerland \\
  ~$^2$ Institute for Particle Physics Phenomenology, Department of Physics,\\Durham University, Durham DH1 3LE, UK\\
  ~$^3$ Rudolf Peierls Centre for Theoretical Physics, University of Oxford,\\ Oxford OX1 3NP, UK\\
}
\end{center}

\vspace{0.1cm}

\begin{center}
  {\bf \large Abstract}
\end{center}
  We present a new release of the {\tt QCDLoop} library based on a
  modern object-oriented framework. We discuss the available new
  features such as the extension to the complex masses, the
  possibility to perform computations in double and quadruple
  precision simultaneously, and useful caching mechanisms to improve
  the computational speed. We benchmark the performance of the new
  library, and provide practical examples of phenomenological
  implementations by interfacing this new library to Monte Carlo
  programs.

\newpage

\noindent {\Large \textbf{Program Summary}}\\

\noindent {\em Name of the program\/}: {\tt QCDLoop} \\[2mm]
{\em Version\/}: 2.0.0 \\[2mm]
{\em Program obtainable from\/}:
\url{http://cern.ch/qcdloop}
\\[2mm]
{\em Distribution format\/}: compressed tar file from the {\tt GitHub} git repository \\[2mm]
{\em E-mail\/}: {\tt stefano.carrazza@cern.ch}, {\tt keith.ellis@durham.ac.uk} \, {\tt giulia.zanderighi@cern.ch} \\[2mm]
{\em License\/}: GNU Public License GPLv3 \\[2mm]
{\em Computers\/}: all \\[2mm]
{\em Operating systems\/}: all with a {\tt c++11} compliant compiler with {\tt quadmath} support, see Sect.~\ref{sec-manual}. \\[2mm]
{\em Program language\/}: {\tt c/c++}, {\tt fortran~77/90}, and {\tt python} \\[2mm]
{\em Memory required to execute\/}:  $\lesssim$ 2 MB \\[2mm]
{\em Other programs called\/}: None \\[2mm]
{\em External files needed\/}: None \\[2mm]
{\em Number of bytes in distributed program, including test data
  etc.\/}: $\sim 1.0$~MB\\[2mm]
\\[2mm]
{\em Nature of the physical problem\/}: Computation of one-loop scalar integrals
\\[2mm]
{\em Solution Method\/}: Numerical evaluation of one-loop scalar integrals such as tadpole, bubble, triangle and box through analytic expressions.
\\[2mm]
{\em Typical running time\/}: detailed performance benchmark presented in Sect.~\ref{sec-benchmarks}

\clearpage

\tableofcontents

\clearpage

\section{Introduction}
\label{sec:intro}

The requirements of precision physics at the LHC and future
experiments demand high precision theoretical predictions. 
In this context perturbative expansions in the coupling constant play
a prominent role.  The field of next-to-leading order (NLO) QCD
corrections has undergone a revolution in the last 10-15 years, see
e.g.~Refs.~\cite{Ellis:2011cr,Andersen:2014efa} and references
therein.  This revolution resulted in computational tools that allow
one to obtain NLO results for generic processes in a semi- or fully
automated
way~\cite{Berger:2008sj,Cullen:2011ac,Cascioli:2011va,Alwall:2014hca}. One
longstanding bottleneck in NLO calculations has been the computation
of virtual corrections. Recently, it was understood how to use
algebraic methods to write any virtual amplitude as a product of
coefficients (that can be computed essentially as products of tree
level amplitudes) and of one-loop scalar master integrals.
Still, for complicated final states often several CPU years are required to obtain distributions that are smooth both at low and high momentum scales. Practically, this means that one needs to run codes for several days on computer farms. Obviously any improvement in the performance of these tools would be welcome. 

In more recent years, a NNLO (next-to-next-to-leading) revolution has
also started, and now almost all $2\to2$ Standard Model processes are
known to this accuracy. One of the ingredients required to achieve
NNLO accuracy for $pp \to X$ is a pure NLO prediction $pp \to X
+1$~parton in the kinematic configuration where the parton becomes
unresolved. Hence, one loop scalar integrals are a crucial ingredient
for both NLO and NNLO calculations, and a fast computation of these
integrals, that remains stable also in kinematic regions where
external partons become soft or collinear, is required.  In the
original paper by two of us (Ellis and Zanderighi~\cite{Ellis:2007qk})
an algorithm was provided to calculate all the divergent one-loop
integrals.  The algorithm proceeds by defining a basis set of
divergent integrals, some of which were available in the literature
prior to ref.~\cite{Ellis:2007qk}, and some of which were calculated
{\it ab initio}.  The results for all the divergent integrals in the
basis set were given in ref.~\cite{Ellis:2007qk} and they were
implemented in a fortran code, dubbed {\tt QCDLoop}. For finite
triangle- and box-integrals {\tt QCDLoop} relied on {\tt
  ff}~\cite{vanOldenborgh:1990yc}. One-loop scalar integrals have been
implemented also in a number of other packages: {\tt
  LoopTools}~\cite{Hahn:2006qw}, {\tt
  OneLoop}~\cite{vanHameren:2010cp} and {\tt
  Collier}~\cite{Denner:2016kdg}.  In the case of unstable particles,
calculations are often performed in the complex mass
scheme~\cite{Denner:2006ic}. So far, the {\tt QCDLoop} library was
limited to real masses in the propagators. Here we present an
extension of the package to deal with complex masses.
%
%high performance computational tools...
More generally, the aim of this paper is to present {\tt
  QCDLoop~2.0}\footnote{In the following sections the label ``{\tt
    QCDLoop}'' refers to the new library}, a new library written in
{\tt c++} and based on the {\tt QCDLoop~1.96} formalism documented in
Ref.~\cite{Ellis:2007qk}. This new framework includes, new features
such as the extension to complex masses, the possibility to switch
precision from double to quadruple precision on the fly. Furthermore
this new framework provides an abstract object-oriented inheritance
mechanism which simplifies the implementation of caching algorithms
which is useful when high performance is required.

The outline of this paper is the following. In
Section~\ref{sec-formalism} we present a short summary of the analytic
expressions and relative diagrams implemented in {\tt QCDLoop}. In
Section~\ref{sec-manual} we describe the library structure and we
introduce the main functionalities of {\tt QCDLoop} and describe the
standard user interface. In Section~\ref{sec-benchmarks} we present a
detailed performance benchmark followed by results obtained with the
integration of {\tt QCDLoop} in a few public Monte Carlo simulation
codes. Finally, in Section~\ref{sec-conclusion} we summarize the
features and advantages of the new {\tt QCDLoop} package.

\section{One-loop scalar integral formalism}
\label{sec-formalism}

\begin{figure}[t]
\centering
\includegraphics[scale=0.25]{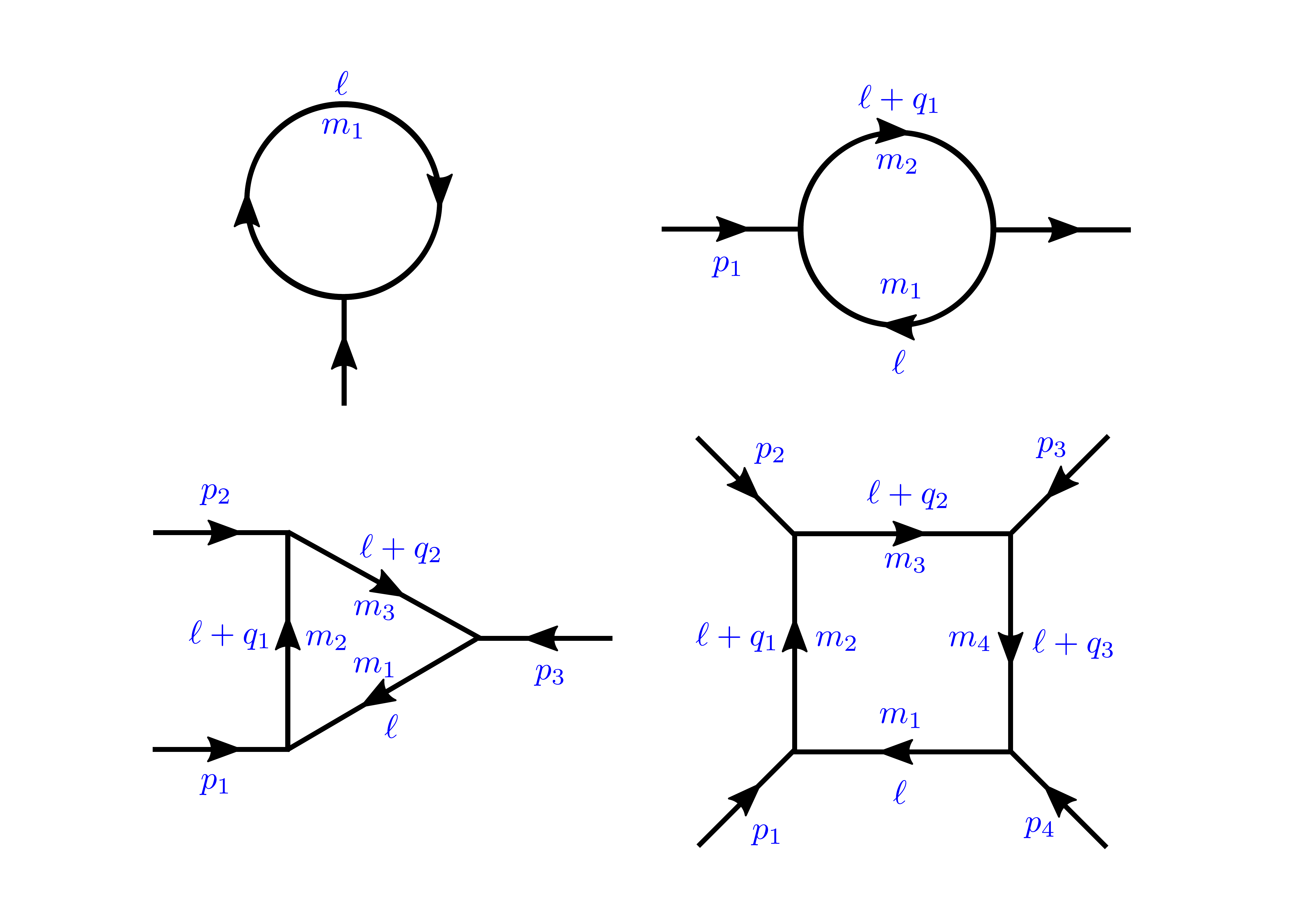}
\caption{\small The notation for the one-loop tadpole, bubble,
  triangle and box integrals.}
\label{fig:diagrams}
\end{figure}

The {\tt QCDLoop} library provides the numerical evaluation of
one-loop scalar integrals such as tadpole, bubble, triangle and box
through analytic expressions. This set of integrals constitutes a basis 
for one-loop scalar integrals. 
In Fig.~\ref{fig:diagrams} we provide a graphical representation for
the definition of the following integrals:
\begin{itemize}
\item Tadpole:
  \begin{equation}
    I_1^{D}(m_1^2) = \frac{\mu^{4-D}}{i\pi^{\frac{D}{2}}r_\Gamma} \int d^Dl \frac{1}{\left( l^2-m_1^2 + i \epsilon \right)},
  \end{equation}
\item Bubble:
  \begin{equation}
    I_2^{D}(p_1;m_1^2,m_2^2) = \frac{\mu^{4-D}}{i\pi^{\frac{D}{2}}r_\Gamma} \int d^Dl \frac{1}{\left( l^2-m_1^2+i\epsilon \right) \left( (l+q_1)^2 - m_2^2 + i\epsilon \right)},
  \end{equation}
\item Triangle:
  \begin{align}
   & I_3^{D}(p_1^2,p_2^2,p_3^2;m_1^2,m_2^2,m_3^2) = \frac{\mu^{4-D}}{i\pi^{\frac{D}{2}}r_\Gamma} \nonumber \\ 
   & \times \int d^Dl \frac{1}{\left( l^2-m_1^2+i\epsilon \right) \left( (l+q_1)^2 - m_2^2 + i\epsilon \right) \left( (l+q_2)^2 - m_3^2 + i\epsilon \right)},
  \end{align}
\item Box:
  \begin{align}
    & I_4^{D}(p_1^2,p_2^2,p_3^2,p_4^2;s_{12},s_{23};m_1^2,m_2^2,m_3^2,m_4^2) = \frac{\mu^{4-D}}{i\pi^{\frac{D}{2}}r_\Gamma} \nonumber \\
    & \times \int d^Dl \frac{1}{\left( l^2-m_1^2+i\epsilon \right) \left( (l+q_1)^2 - m_2^2 + i\epsilon \right) \left( (l+q_2)^2 - m_3^2 + i\epsilon \right) \left( (l+q_3)^2 - m_4^2 + i\epsilon \right)},
  \end{align}
\end{itemize}
where $q_n \equiv \sum_{i=1}^n p_i$
%, $q_0=0$ 
and
$s_{ij}=(p_i+p_j)^2$. The above expressions are in the Bjorken-Drell
metric so that $l^2=l_0^2-l_1^2-l_2^2-l_3^2$. In this paper we
consider momenta to be real, but the masses to be either real or
complex. Near four dimensions we use $D=4-2\epsilon$ and $\mu$ is a
scale introduced so that the integrals preserve their natural
dimensions, despite excursions away from $D=4$. We have also removed
the overall constant term which occurs in $D-$dimensional integrals
\begin{equation}
  r_\Gamma \equiv \frac{\Gamma^2(1-\epsilon)\Gamma(1+\epsilon)}{\Gamma(1-2\epsilon)} = \frac{1}{\Gamma(1-\epsilon)} + \mathcal{O}(\epsilon^3) = 1 - \epsilon \gamma + \epsilon^2 \left[ \frac{\gamma^2}{2} - \frac{\pi^2}{12}\right] + \mathcal{O}(\epsilon^3).
\end{equation}

The explicit expressions for all the divergent integrals in 
{\tt QCDLoop 2.0} are presented in detail in the original 
paper of Ellis and Zanderighi~\cite{Ellis:2007qk}. 
As noted in that paper, some of the results for the divergent integrals 
were new, but many of them were not. We refer the reader to 
Ref.~\cite{Ellis:2007qk} for details and the appropriate references. 
These expressions have been adjusted for performance optimization and the
proper analytical continuation has been performed to deal with complex
masses. The finite integrals for the bubble topology are taken from
Refs.~\cite{Denner:2005nn,Denner:1991kt}.
Finite integrals for the triangle topology with real masses are
obtained from Refs.~\cite{'tHooft:1978xw,vanOldenborgh:1989wn}
following the {\tt LoopTools} implementation. Expressions for complex
masses can be derived from
Refs.~\cite{Denner:1991qq,Denner:2005nn,'tHooft:1978xw,vanOldenborgh:1989wn}. We
found that the expressions of Ref.~\cite{Denner:1991qq}, as
implemented in {\tt OneLoop}, had the best performance and hence
this was the approach that we followed in our implementation.
Finally, the finite box integrals are based on
Ref.~\cite{Denner:1991qq}, following the {\tt LoopTools} implementation.
We invite the reader to examine the {\tt QCDLoop 2.0} source code and 
documentation for further details.

%%%%%%%%%%%%%%%%%%%%%%%%%%%%%%%%%%%%%%%%%%%%%%%%%%%%%%%%%%
%%%%%%%%%%%%%%%%%%%%%%%%%%%%%%%%

\section{{\tt QCDLoop} library documentation}
\label{sec-manual}

In this section we present the user manual for the {\tt QCDLoop}
library. First of all, we discuss how to download and install {\tt
  QCDLoop}. After that, we illustrate the new framework design in {\tt
  c++}, explaining how objects are organized and how to run a simple
program. We conclude the discussion by presenting the available
caching mechanisms and the {\tt fortran} and {\tt python} wrappers.

\subsection{Download and installation}
\label{sec:install}

The {\tt QCDLoop} library is available from the website:
\begin{center}
{\bf \url{http://cern.ch/qcdloop}~}
\end{center}

The installation of the {\tt QCDLoop} library can be easily performed
using the standard {\tt autotools} sequence:
\begin{lstlisting}
   ./configure
   make
   make install
\end{lstlisting}
which automatically installs {\tt QCDLoop} in {\tt /usr/local/}. 
Note that the {\tt QCDLoop} library requires a {\tt c++11} compliant
compiler with {\tt quadmath} support, such as {\tt g++ 4.7\footnote{\url{https://gcc.gnu.org/}}, 
icpc 13\footnote{\url{https://software.intel.com/}}}
and or more recent versions of these compilers. 
The configure script will check for these and
other system requirements before building the {\tt makefiles}.
In order to use a different installation path one can use the option:
\begin{lstlisting}
   ./configure --prefix=<path to the installation folder>
\end{lstlisting}
In this case, the {\tt QCDLoop} installation path should be included
to the environment variables {\tt PATH} and {\tt LD\_LIBRARY\_PATH},
adding to the local {\tt .bashrc} file (or {\tt .profile} file on Mac)
the string:
\begin{lstlisting}
  export PATH=$PATH:<installation folder>/bin
  export LD_LIBRARY_PATH=$LD_LIBRARY_PATH:<installation folder>/lib
\end{lstlisting}

If the system provides more than one {\tt c++} compiler we suggest to
set the preferable choice when running {\tt configure}:
\begin{lstlisting}
   ./configure CXX=<compiler name / path>
\end{lstlisting}

Finally, this package provides a {\tt qcdloop-config} tool which
simplifies the usage of the library when linking and compiling with
user's codes. We provide the following flags: {\tt --help}: shows the
help message; {\tt --prefix}: shows the installation prefix; {\tt
  --incdir}: shows the path to the qcdloop header directory; {\tt
  --libdir}: shows the path to the qcdloop library directory; {\tt
  --cppflags}: gets compiler flags for use with the C preprocessor
stage of {\tt c++} compilation; {\tt --ldflags}: gets compiler flags
for use with the linker stage of any compilation; {\tt --version}:
returns qcdloop release version number

\subsection{The framework design}

The development of a new framework for {\tt QCDLoop} is motivated
by the following needs:
\begin{itemize}
\item generalization of the code to support complex masses;
\item provision of the ability to provide double and quadruple precision results simultaneously;
\item the removal of code dependencies such as the {\tt ff} library~\cite{vanOldenborgh:1990yc};
\item improvement of performance by implementing more sophisticated {\tt LRU} cache algorithms;
\item provision of a modern framework based on an object-oriented language,
  such as {\tt c++}, which simplifies future developments and native
  integration with modern codes.
\end{itemize}

%%%%%%%%%%%%%%%%%%%%%%%%%%%%%%%%%%%%%%%%%%%%%%%%%%%%%%%%%%
\begin{figure}[t]
\centering
\includegraphics[scale=0.5]{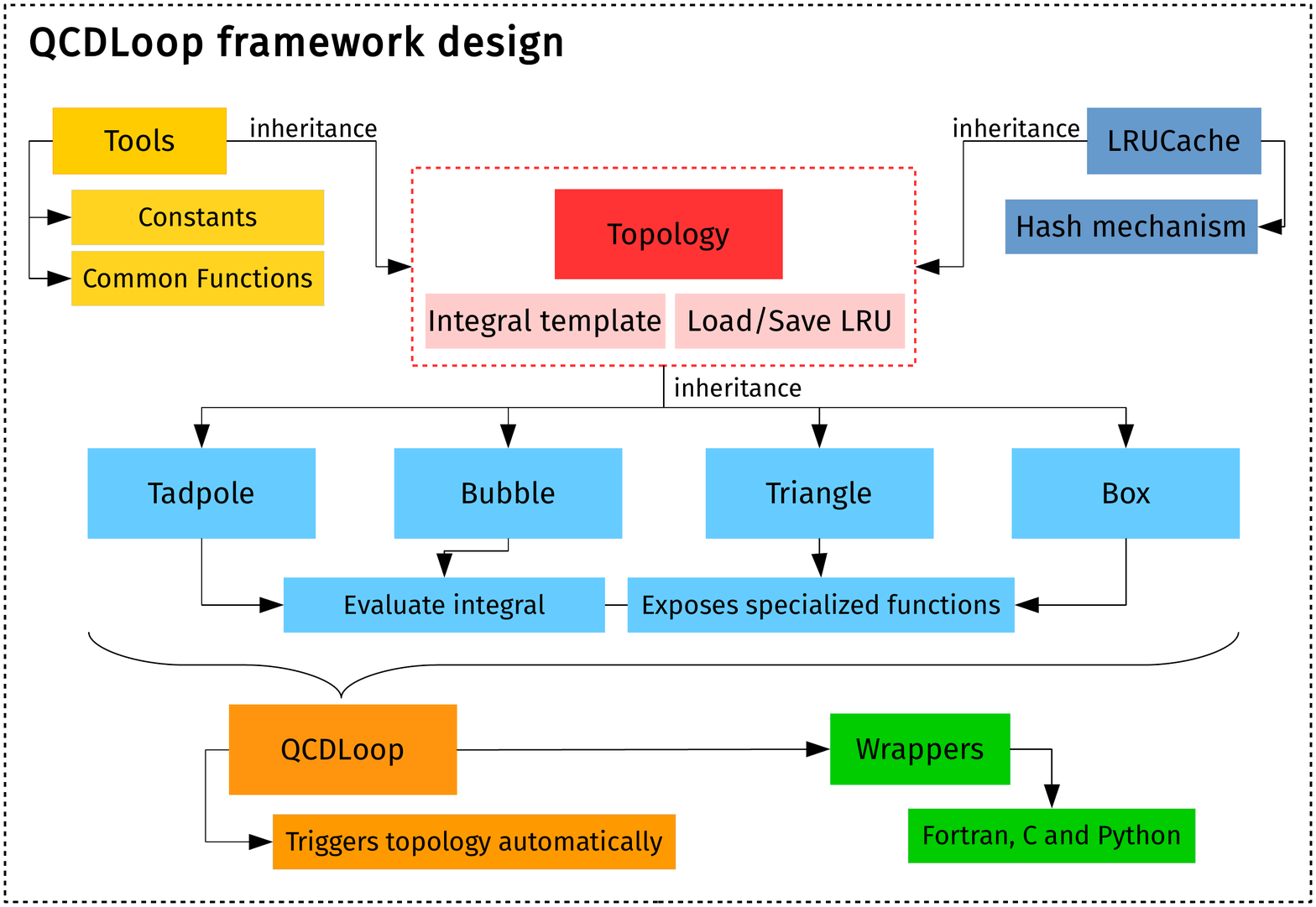}
\caption{\small {\tt QCDLoop} framework design.}
\label{fig:design}
\end{figure}
%%%%%%%%%%%%%%%%%%%%%%%%%%%%%%%%

The new {\tt QCDLoop} framework design is presented schematically in
Fig.~\ref{fig:design}. The core of the code is the {\tt Topology}
class which inherits common methods from {\tt Tools} and {\tt
  LRUCache} classes. {\tt Topology} is a templated class which
provides the public pure virtual {\tt integral} method for the
specialization of the one-loop scalar integrals, together with the
basic mechanism for load and storage of cached results.
The {\tt Tools} templated class initializes real and complex constants
based on the allocated precision type, and it provides a common set of
functions which are shared by the specialization of topologies. It is
important to highlight that this template class reduces to a minimum
the duplication of code when a method or constant is required by several
specializations.
The {\tt LRUCache} class implements a ``latest recent used'' (LRU)
caching algorithm for dynamic caching sizes which are discussed in
detail in Sec.~\ref{sec:cache}.

The specialization of the one-loop scalar integrals are implemented in
four classes, one for each topology: {\tt Tadpole}, {\tt Bubble}, {\tt
  Triangle} and {\tt Box}. These classes implement the public pure
virtual {\tt integral} method from {\tt Topology}. For each topology
the {\tt integral} method triggers automatically the kinematics and
selects the appropriate function call, when possible it also provides
the public methods for the direct computation of the specific
kinematics. A detailed summary of the function calls and the
corresponding class methods are shown in Table~\ref{tab:topo} using
the notation from Ref.~\cite{Ellis:2007qk}.

On top of the specialized topologies we provide the {\tt QCDLoop}
class. This is a high-level user interface which detects the topology
by checking the size of the {\tt integral} arguments. This function
verifies the consistency of the input arguments and calls the
respective topology:
\begin{lstlisting}
/*!
 * Standard arguments to retrieve one-loop scalar integrals.
 * output:
 *   res: vector of dim(3) containing the coefficients in the Laurent series
 *        res[0]: finite part (1/eps^0)
 *        res[1]: single pole (1/eps^1)
 *        res[2]: double pole (1/eps^2)
 * input:
 *   mu2: is the square of the scale mu
 *   m: array containing the squares of the masses of the internal lines
 *   p: array containing the four-momentum squared of the external lines
 */
ql::QCDLoop<TOutput,TMass,TScale>::integral(vector<TOutput> &res,
                                            TScale const& mu2,
                                            vector<TMass> const& m,
                                            vector<TScale> const& p) const;
\end{lstlisting}
The trigger mechanism employed by the {\tt QCDLoop} class is the
safest and simplest way to access all the library functionalities,
however when maximum performance is required the user is invited to
allocate the specific topology in order to remove the overhead due to
the triggering procedure.

We conclude the description of the framework design by highlighting
the availability of {\tt c++} wrappers to {\tt fortran}, {\tt c} and
{\tt python}, further details about these interfaces are presented in
Sect.~\ref{sec:wrappers}.
The native {\tt c++} interface of the {\tt QCDLoop} library
is thread-safe only when the caching algorithms are switched off. It
is particularly important to highlight that the {\tt fortran} wrapper
is not thread-safe by construction.

Further details about the code are fully documented using the {\tt
  doxygen}\footnote{\url{www.doxygen.org}} syntax. The respective
documentation is located in the {\tt qcdloop/doc} folder.

{\small
\begin{table}
  \begin{centering}

\begin{tabular}{|l|l|l|}
\hline 
Function & Short name & Method specialization \tabularnewline
\hline 
\hline 
$I_1^D(m^2)$ & {\tt Tadpole} & {\tt TadPole<>::integral()} \tabularnewline
\hline
\hline
$I_2^D(s;m_1^2,m_2^2)$ & {\tt Bubble finite (BB0)} & {\tt Bubble<>::BB0()} \tabularnewline
\hline 
$I_2^D(m^2;0,m^2)$ &  {\tt Bubble BB1} & {\tt Bubble<>::BB1()} \tabularnewline
\hline
$I_2^D(0;0,m^2)$ &  {\tt Bubble BB2} & {\tt Bubble<>::BB2()} \tabularnewline
\hline
$I_2^D(s;0,0)$ &  {\tt Bubble BB3} & {\tt Bubble<>::BB3()} \tabularnewline
\hline
$I_2^D(s;0,m^2)$ &  {\tt Bubble BB4} & {\tt Bubble<>::BB4()} \tabularnewline
\hline
$I_2^D(0;m_1^2,m_2^2)$ &  {\tt Bubble BB5} & {\tt Bubble<>::BB5()} \tabularnewline
\hline
\hline
$I_3^D(0,0,0;m_1^2,m_2^2,m_3^2)$ &  {\tt Triangle finite (TIN0)} & {\tt Triangle<>::TIN0()} \tabularnewline
\hline
$I_3^D(0,0,p_3^2;m_1^2,m_2^2,m_3^2)$ &  {\tt Triangle finite (TIN1)} & {\tt Triangle<>::TIN1()} \tabularnewline
\hline
$I_3^D(0,p_2^2,p_3^2;m_1^2,m_2^2,m_3^2)$ &  {\tt Triangle finite (TIN2)} & {\tt Triangle<>::TIN2()} \tabularnewline
\hline
$I_3^D(p_1^2,p_2^2,p_3^2;m_1^2,m_2^2,m_3^2)$ &  {\tt Triangle finite (TIN3)} & {\tt Triangle<>::TIN3()} \tabularnewline
\hline
$I_3^D(0,0,p_3^2;0,0,0)$ &  {\tt Triangle T1} & {\tt Triangle<>::T1()} \tabularnewline
\hline
$I_3^D(0,p_2^2,p_3^2;0,0,0)$ &  {\tt Triangle T2} & {\tt Triangle<>::T2()} \tabularnewline
\hline
$I_3^D(0,p_2^2,p_3^2;0,0,m^2)$ &  {\tt Triangle T3} & {\tt Triangle<>::T3()} \tabularnewline
\hline
$I_3^D(0,p_2^2,m^2;0,0,m^2)$ &  {\tt Triangle T4} & {\tt Triangle<>::T4()} \tabularnewline
\hline
$I_3^D(0,m^2,m^2;0,0,m^2)$ &  {\tt Triangle T5} & {\tt Triangle<>::T5()} \tabularnewline
\hline
$I_3^D(m_2^2,s,m_3^2;0,m_2^2,m_3^2)$ &  {\tt Triangle T6} & {\tt Triangle<>::T6()} \tabularnewline
\hline
\hline
$I_4^D(p_1^2,p_2^2,p_3^2,p_4^2;s_{12},s_{23};0,0,0,0)$ &  {\tt Box finite (BIN0)} & {\tt Box<>::BIN0()} \tabularnewline
\hline
$I_4^D(p_1^2,p_2^2,p_3^2,p_4^2;s_{12},s_{23};0,0,0,m_4^2)$ &  {\tt Box finite (BIN1)} & {\tt Box<>::BIN1()} \tabularnewline
\hline
$I_4^D(p_1^2,p_2^2,p_3^2,p_4^2;s_{12},s_{23};0,0,m_3^2,m_4^2)$ &  {\tt Box finite (BIN2)} & {\tt Box<>::BIN2()} \tabularnewline
\hline
$I_4^D(p_1^2,p_2^2,p_3^2,p_4^2;s_{12},s_{23};0,m_2^2,m_3^2,m_4^2)$ &  {\tt Box finite (BIN3)} & {\tt Box<>::BIN3()} \tabularnewline
\hline
$I_4^D(p_1^2,p_2^2,p_3^2,p_4^2;s_{12},s_{23};m_1^2,m_2^2,m_3^2,m_4^2)$ &  {\tt Box finite (BIN4)} & {\tt Box<>::BIN4()} \tabularnewline
\hline

$I_4^D(0,0,0,0;s_{12},s_{23};0,0,0,0)$ &  {\tt Box B1} & {\tt Box<>::B1()} \tabularnewline
\hline
$I_4^D(0,0,0,p_4^2;s_{12},s_{23};0,0,0,0)$ &  {\tt Box B2} & {\tt Box<>::B2()} \tabularnewline
\hline
$I_4^D(0,p_2^2,0,p_4^2;s_{12},s_{23};0,0,0,0)$ &  {\tt Box B3} & {\tt Box<>::B3()} \tabularnewline
\hline
$I_4^D(0,0,p_3^2,p_4^2;s_{12},s_{23};0,0,0,0)$ &  {\tt Box B4} & {\tt Box<>::B4()} \tabularnewline
\hline
$I_4^D(0,p_2^2,p_3^2,p_4^2;s_{12},s_{23};0,0,0,0)$ &  {\tt Box B5} & {\tt Box<>::B5()} \tabularnewline
\hline
$I_4^D(0,0,m^2,m^2;s_{12},s_{23};0,0,0,m^2)$ &  {\tt Box B6} & {\tt Box<>::B6()} \tabularnewline
\hline
$I_4^D(0,0,m^2,p_4^2;s_{12},s_{23};0,0,0,m^2)$ &  {\tt Box B7} & {\tt Box<>::B7()} \tabularnewline
\hline
$I_4^D(0,0,p_3^2,p_4^2;s_{12},s_{23};0,0,0,m^2)$ &  {\tt Box B8} & {\tt Box<>::B8()} \tabularnewline
\hline
$I_4^D(0,p_2^2,p_3^2,m^2;s_{12},s_{23};0,0,0,m^2)$ &  {\tt Box B9} & {\tt Box<>::B9()} \tabularnewline
\hline
$I_4^D(0,p_2^2,p_3^2,p_4^2;s_{12},s_{23};0,0,0,m^2)$ &  {\tt Box B10} & {\tt Box<>::B10()} \tabularnewline
\hline
$I_4^D(0,m_3^2,p_3^2,m_4^2;s_{12},s_{23};0,0,m_3^2,m_4^2)$ &  {\tt Box B11} & {\tt Box<>::B11()} \tabularnewline
\hline
$I_4^D(0,m_3^2,p_3^2,p_4^2;s_{12},s_{23};0,0,m_3^2,m_4^2)$ &  {\tt Box B12} & {\tt Box<>::B12()} \tabularnewline
\hline
$I_4^D(0,p_2^2,p_3^2,p_4^2;s_{12},s_{23};0,0,m_3^2,m_4^2)$ &  {\tt Box B13} & {\tt Box<>::B13()} \tabularnewline
\hline
$I_4^D(m_2^2,m_2^2,m_4^2,m_4^2;s_{12},s_{23};0,m_2^2,0,m_4^2)$ &  {\tt Box B14} & {\tt Box<>::B14()} \tabularnewline
\hline
$I_4^D(m_2^2,p_2^2,p_3^2,m_4^2;s_{12},s_{23};0,m_2^2,0,m_4^2)$ &  {\tt Box B15} & {\tt Box<>::B15()} \tabularnewline
\hline
$I_4^D(m_2^2,p_2^2,p_3^2,m_4^2;s_{12},s_{23};0,m_2^2,m_3^2,m_4^2)$ &  {\tt Box B16} & {\tt Box<>::B16()} \tabularnewline
\hline

\end{tabular}
\par\end{centering}
\caption{\label{tab:topo}Summary of the function calls and the
  corresponding class methods in {\tt QCDLoop} following the notation
  from Ref.~\cite{Ellis:2007qk}. The second column refers to the
  labels used in Figs.~\ref{fig:comparison}
  and \ref{fig:comparison2}. Note that finite configurations do not
  require $D$ dimension.}

\end{table}
}

\subsubsection{The {\tt namespace ql}: types, typedefs and templates}

From a technical point of view all objects of the {\tt QCDLoop}
library are implemented in the {\tt ql} namespace. In this namespace
we define also aliases for double and quadruple precision real and
complex types so that the primitive {\tt ql} types are {\tt double},
{\tt qdouble}, {\tt complex} and {\tt qcomplex}:
\begin{lstlisting}
namespace ql
{
  typedef __float128 qdouble;           // quadruple precision real type
  typedef __complex128 qcomplex;        // quadruple precision complex type
  typedef std::complex<double> complex; // double precision complex type
}
\end{lstlisting}
The quadruple real and complex types are standard {\tt quadmath}
objects, and the double complex type corresponds to the {\tt std::complex} type. 
Specialized mathematical operations are
implemented for each type in the inline header {\tt qcdloop/math.h}.

In order to allocate simultaneously double and quadruple precision
objects, all classes presented in Fig.~\ref{fig:design} are templated
with three typenames: {\tt TOutput} the output type, {\tt TMass} the
mass type and {\tt TScale} the scale and momenta type. The accepted
types for each typename is listed below:
\begin{lstlisting}
 // Typenames and possible combinations (columns)
 typename TOutput -> complex | qcomplex | complex | qcomplex
 typename TMass   -> double  | qdouble  | complex | qcomplex
 typename TScale  -> double  | qdouble  | double  | qdouble
\end{lstlisting}
Template classes are locked to these combinations. The compiler
prevents the allocation of wrong combinations at compilation time.
Further extensions for the typenames are possible if required.

\subsubsection{Code examples}
\label{sec:code}

A simple example of code usage in {\tt c++} is presented in the code
snippet given below. The code shows how to compute a tadpole double
precision integral first by using the {\tt QCDLoop} trigger and then
the direct allocation of the {\tt TadPole} class. A similar example is
then illustrated for the quadruple precision and complex mass
calculation, see comments in the code. In order to change the topology
it is sufficient to modify the content and size of the squared momenta
and mass vectors and the initialization of the specific topology class
if a direct computation is desired.
Further examples are available and are build at the compilation time
in the {\tt examples/} folder.

\begin{lstlisting}
#include <qcdloop/qcdloop.h>
using namespace ql;

int main() {
  // double precision variables
  double mu2 = ql::Pow(1.7,2);
  std::vector<double> p = {};
  std::vector<double> m = {5.0};
  std::vector<complex> res(3);

  // Trigger example - Tadpole double precision with real mass
  ql::QCDLoop<complex,double,double> auto_trigger;
  auto_trigger.integral(res, mu2, m, p);

  // Tadpole direct call - double precision with real mass
  ql::TadPole<complex,double,double> tp;
  tp.integral(res, mu2, m, p);

  // quadruple precision and complex mass variables
  qdouble mu2q = ql::Pow(1.7q,2);
  std::vector<qdouble> pq = {};
  std::vector<qcomplex> mq = { {5.0q,-1.0q} };
  std::vector<qcomplex> resq(3);

  // Trigger example - Tadpole quadruple precision with complex mass
  ql::QCDLoop<qcomplex,qcomplex,qdouble> auto_trigger_q;
  auto_trigger_q.integral(resq, mu2q, mq, pq);

  // Tadpole direct call - quadruple precision with complex mass
  ql::TadPole<qcomplex,qcomplex,qdouble> tpq;
  tpq.integral(resq, mu2q, mq, pq);
  
  return 0; }
\end{lstlisting}

\subsubsection{Caching mechanisms}
\label{sec:cache}

We provide two caching algorithms for fast retrieval of previously
computed one-loop scalar integrals. By default the {\tt Topology} class
implements and allocates a ``last-used'' LU cache, with dimension
$N=1$, where only the last computed result is stored. Such a caching
mechanism is similar to the previous {\tt QCDLoop 1.96} version,
however a faster argument comparison algorithm parser is
employed. Note that this approach is the fastest method when using a
small cache with $N=1$.

Due to the practical limitations of this approach we implement a
dynamic size ``last-recent-used'' (LRU) algorithm in the {\tt
  LRUCache} class. Such approach provides a simple and fast method to
store the last $N$ computed integrals, where $N$ is chosen by the
user. The algorithm first computes a key associated with the integral
arguments by using the Murmur hash algorithm from the {\tt std::Hash}
function available from the {\tt c++} standard library. We have
verified explicitly that the rate of hash collisions is negligible in
the context of one-loop scalar integral computations. Secondly, the
result of the integral is stored in an unordered map so that the
searching mechanism is based on a single key search. Performance
results are presented and discussed in detail in
Sec.~\ref{sec-performance}.

In order to activate the different caching algorithms the user should
call the {\tt setCacheSize(int const\& size)} method which is
available from all inherited classes from {\tt Topology} and {\tt
  QCDLoop}. The code automatically selects the appropriate caching
algorithm based on the {\tt size} parameter:
\begin{lstlisting}
  ql::TadPole<complex,double,double> tp; // default cache size N = 1
  tp.setCacheSize(10); // sets the cache to N=10
  // perform calculation...
\end{lstlisting}
Note the possibility to switch off the caching algorithm by setting
{\tt size = 0}.

\subsubsection{{\tt Fortran} and {\tt python} wrappers}
\label{sec:wrappers}

The {\tt QCDLoop} library provides wrappers to {\tt fortran (77/90)}
and {\tt c} based on the same syntax of {\tt QCDLoop 1.96}
in~\cite{Ellis:2007qk}. Table~\ref{tab:wrapper} lists the available
functions for different topologies and argument types. We enlarge the
previous {\tt qlI$j$} ($j=1,2,3,4$) syntax with extra functions
identified by new prefixes: {\tt qlI$j$c} computes integrals in double
precision and complex masses, {\tt qlI$j$q} computes integrals in
quadruple precision and real masses, {\tt qlI$j$qc} computes integrals
in quadruple precision and complex masses. In parallel to these
functions we included the new {\tt qlcachesize(size)} function which
provides the interface to modify the cache size. Further details of
these wrappers are available in the header {\tt qcdloop/wrapper.h}

{\small
\begin{table}
  \begin{centering}
\begin{tabular}{|c|l|c|c|}
\hline 
Integral & {\tt fortran} function & \multicolumn{2}{c|}{Description}\tabularnewline
\hline 
\hline 
- & {\tt qlinit()} & \multicolumn{2}{c|}{initializes the library}\tabularnewline
\hline 
- & {\tt qlcachesize(size)} & \multicolumn{2}{c|}{sets the cache size}\tabularnewline
\hline 
\hline 
\multicolumn{2}{|c|}{} & Precision & Masses\tabularnewline
\hline 
\hline 
\multirow{4}{*}{$I_{1}^{D}$} & {\tt ql1(m1,mu2,ep)} & {\tt double} & {\tt real}\tabularnewline
\cline{2-4} 
 & {\tt ql1c(m1,mu2,ep)} & {\tt double} & {\tt complex}\tabularnewline
\cline{2-4} 
 & {\tt ql1q(m1,mu2,ep)} & {\tt quadruple} & {\tt real}\tabularnewline
\cline{2-4} 
 & {\tt ql1qc(m1,mu2,ep)} & {\tt quadruple} & {\tt complex}\tabularnewline
\hline 
\multirow{4}{*}{$I_{2}^{D}$} & {\tt qlI2(p1,m1,m2,mu2,ep)} & {\tt double} & {\tt real}\tabularnewline
\cline{2-4} 
 & {\tt qlI2c(p1,m1,m2,mu2,ep)} & {\tt double} & {\tt complex} \tabularnewline
\cline{2-4} 
 & {\tt qlI2q(p1,m1,m2,mu2,ep)} & {\tt quadruple} & {\tt real}\tabularnewline
\cline{2-4} 
 & {\tt qlI2qc(p1,m1,m2,mu2,ep)} & {\tt quadruple} & {\tt complex}\tabularnewline
\hline 
\multirow{4}{*}{$I_{3}^{D}$} & {\tt qlI3(p1,p2,p3,m1,m2,m3,mu2,ep)} & {\tt double} & {\tt real}\tabularnewline
\cline{2-4} 
 & {\tt qlI3c(p1,p2,p3,m1,m2,m3,mu2,ep)} & {\tt double} & {\tt complex}\tabularnewline
\cline{2-4} 
 & {\tt qlI3q(p1,p2,p3,m1,m2,m3,mu2,ep)} & {\tt quadruple} & {\tt real}\tabularnewline
\cline{2-4} 
 & {\tt qlI3qc(p1,p2,p3,m1,m2,m3,mu2,ep)} & {\tt quadruple} & {\tt complex}\tabularnewline
\hline 
\multirow{4}{*}{$I_{4}^{D}$} & {\tt qlI4(p1,p2,p3,p4,s12,s23,m1,m2,m3,m4,mu2,ep)} & {\tt double} & {\tt real}\tabularnewline
\cline{2-4} 
 & {\tt qlI4c(p1,p2,p3,p4,s12,s23,m1,m2,m3,m4,mu2,ep)} & {\tt double} & {\tt complex} \tabularnewline
\cline{2-4} 
 & {\tt qlI4q(p1,p2,p3,p4,s12,s23,m1,m2,m3,m4,mu2,ep)} & {\tt quadruple} & {\tt real}\tabularnewline
\cline{2-4} 
 & {\tt qlI4qc(p1,p2,p3,p4,s12,s23,m1,m2,m3,m4,mu2,ep)} & {\tt quadruple} & {\tt complex}\tabularnewline
\hline 
\end{tabular}
\par\end{centering}
\caption{\label{tab:wrapper}{\tt fortran} and {\tt c} wrapper
  functions.}
\end{table}
}

We also provide a basic {\tt python} interface to the library through
{\tt cython} which can be extended by the user easily. In order to
build and install the {\tt python} module for {\tt QCDLoop} the user
should first install the library following the instructions in
Sect.~\ref{sec:install} (and exporting the {\tt PATH} and {\tt
  LD\_LIBRARY\_PATH} environment variables) and then perform the
following operations:
\begin{lstlisting}
  cd pywrap
  python setup.py install
\end{lstlisting}
The last command invokes the {\tt cython} compiler and installs the
module to the system {\tt PYTHONPATH}.

The {\tt python} wrapper contains the {\tt qcdloop.QCDLoop} object
which reflects exactly the class {\tt ql::QCDLoop} from the
library. One can obtain results by querying the python console with:
\begin{lstlisting}
  # TadPole computation in python
  from qcdloop import QCDLoop as ql
  m = [0.5]
  mu2 = 1.7**2
  out = ql.integral(mu2,m)
\end{lstlisting}

\section{Validation and benchmarks}
\label{sec-benchmarks}

In this section we quantify and benchmark the performance of the new
{\tt QCDLoop} library in terms of computational time and then in terms
of phenomenological results.

\subsection{Performance tests}
\label{sec-performance}

All the topologies implemented in {\tt QCDLoop 2.0} have been
validated successfully by direct comparison with {\tt QCDLoop
  1.96}~\cite{Ellis:2007qk} for configurations with real masses and
{\tt OneLoop 3.6}~\cite{vanHameren:2010cp} for complex masses.

The performance benchmark is based on a common setup: all libraries
compiled with {\tt gcc-5.2.1} using {\tt -O2} optimization flags on a
i7-6500U CPU @ 2.50GHz. Kinematical configurations are constructed
before the computation of the scalar integrals in order to avoid
copy-assignment operations during the computation loop. We use the
native language of each library when performing the benchmark in
order to avoid eventual overhead due to wrapper manipulation. We
remove the initialization time of {\tt OneLoop} from the results.

%%%%%%%%%%%%%%%%%%%%%%%%%%%%%%%%%%%%%%%%%%%%%%%%%%%%%%%%%%
\begin{figure}[hp]
\centering
\includegraphics[scale=0.4]{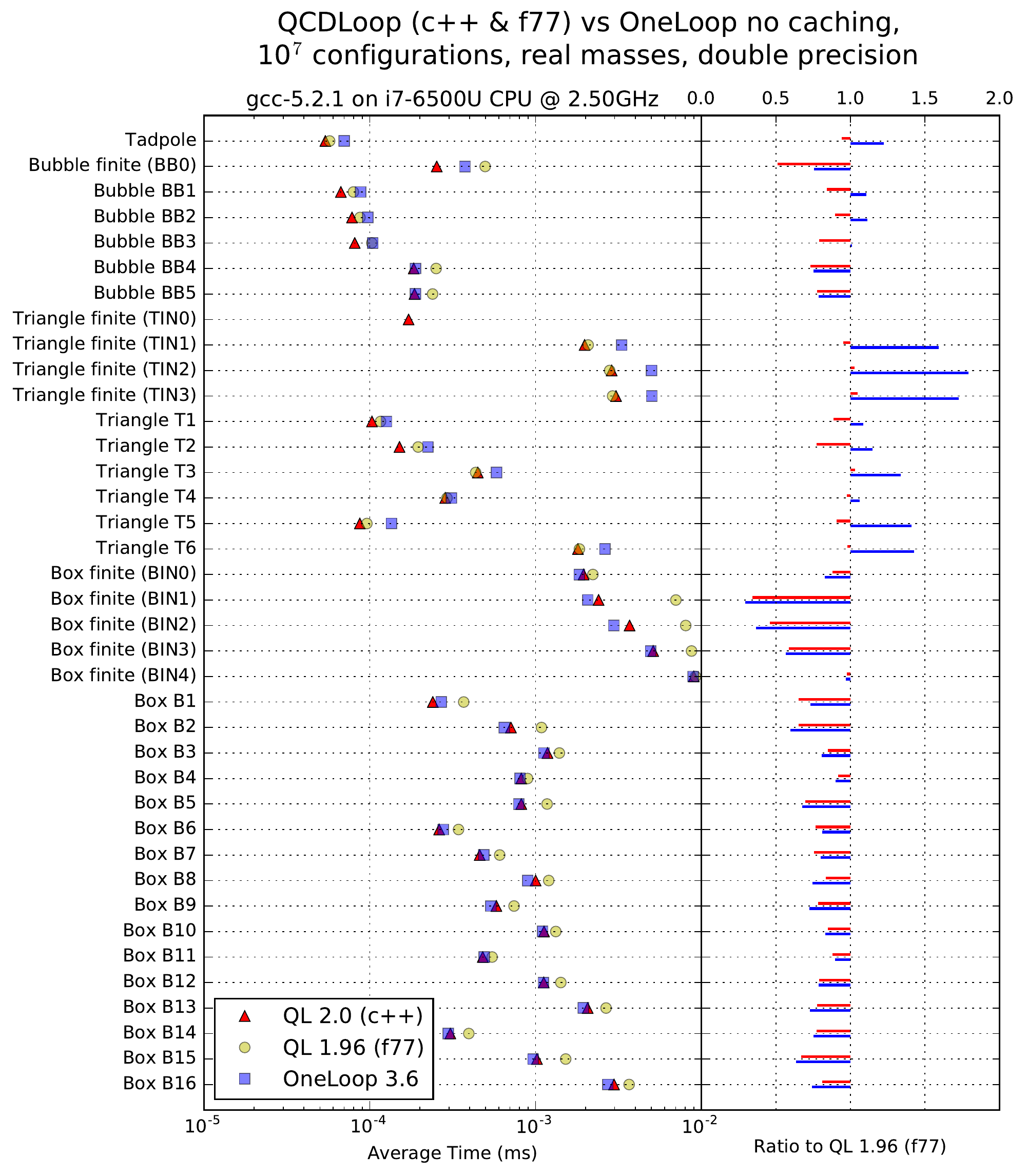}\includegraphics[scale=0.4]{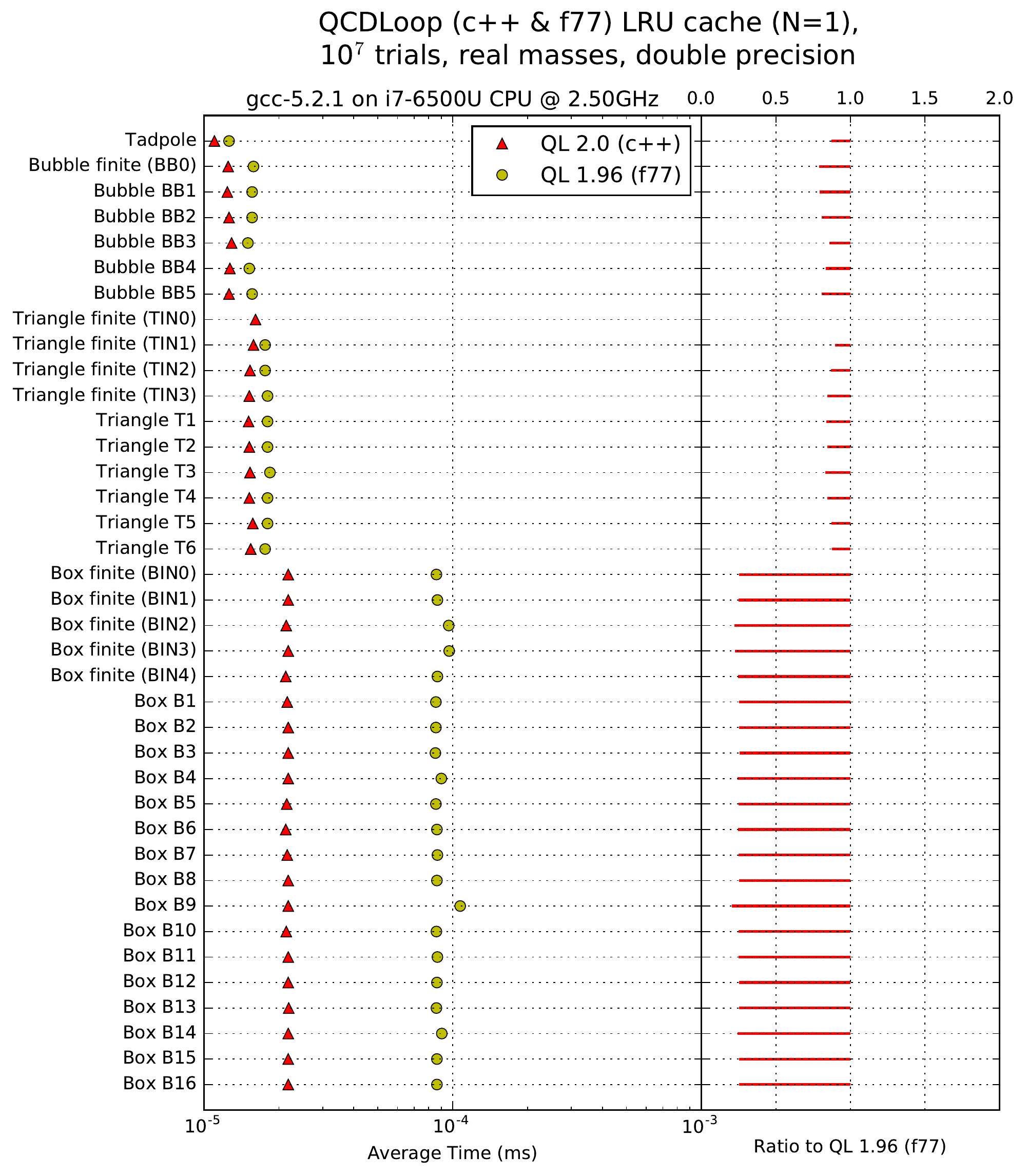}
\includegraphics[scale=0.4]{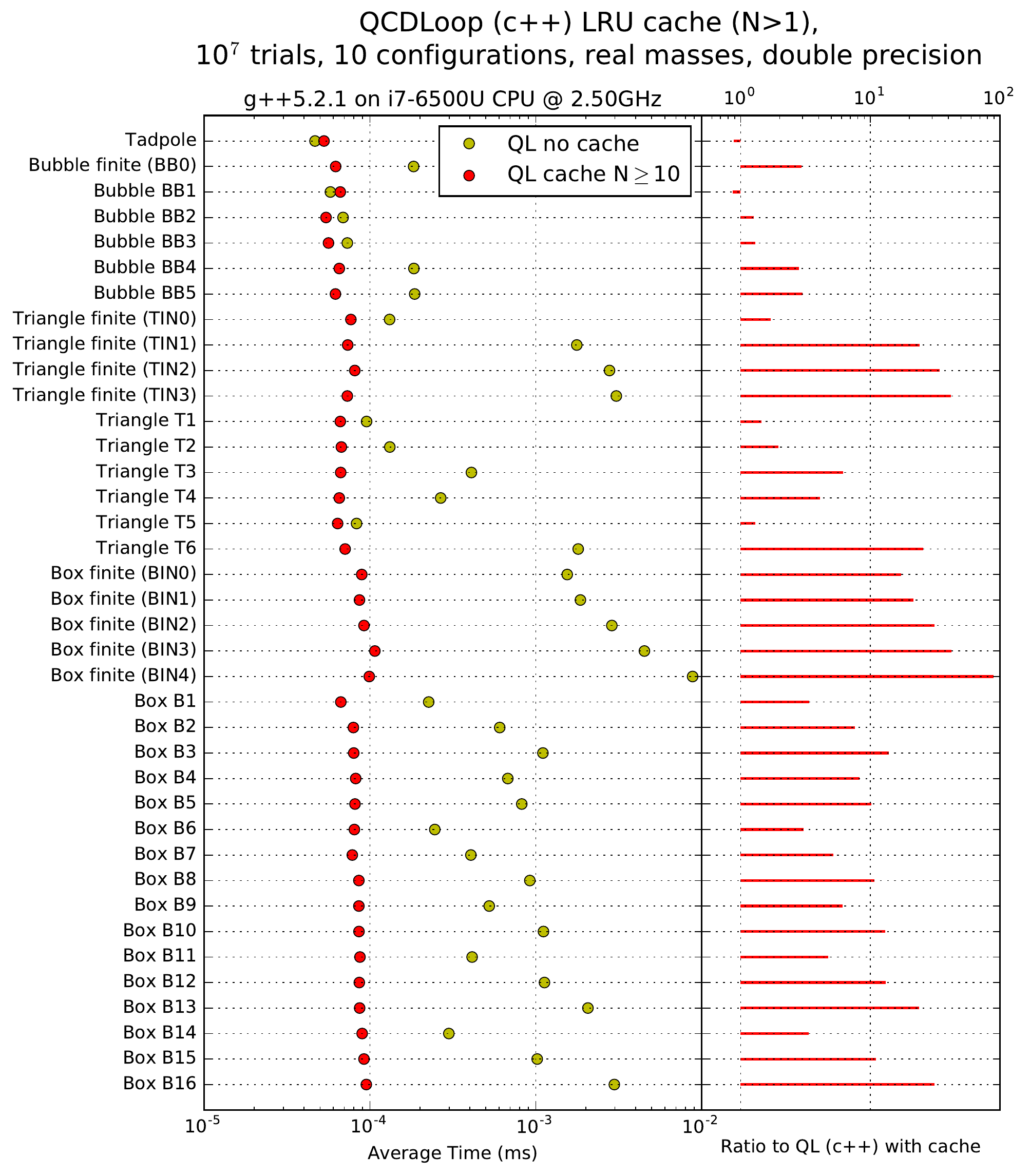}\includegraphics[scale=0.4]{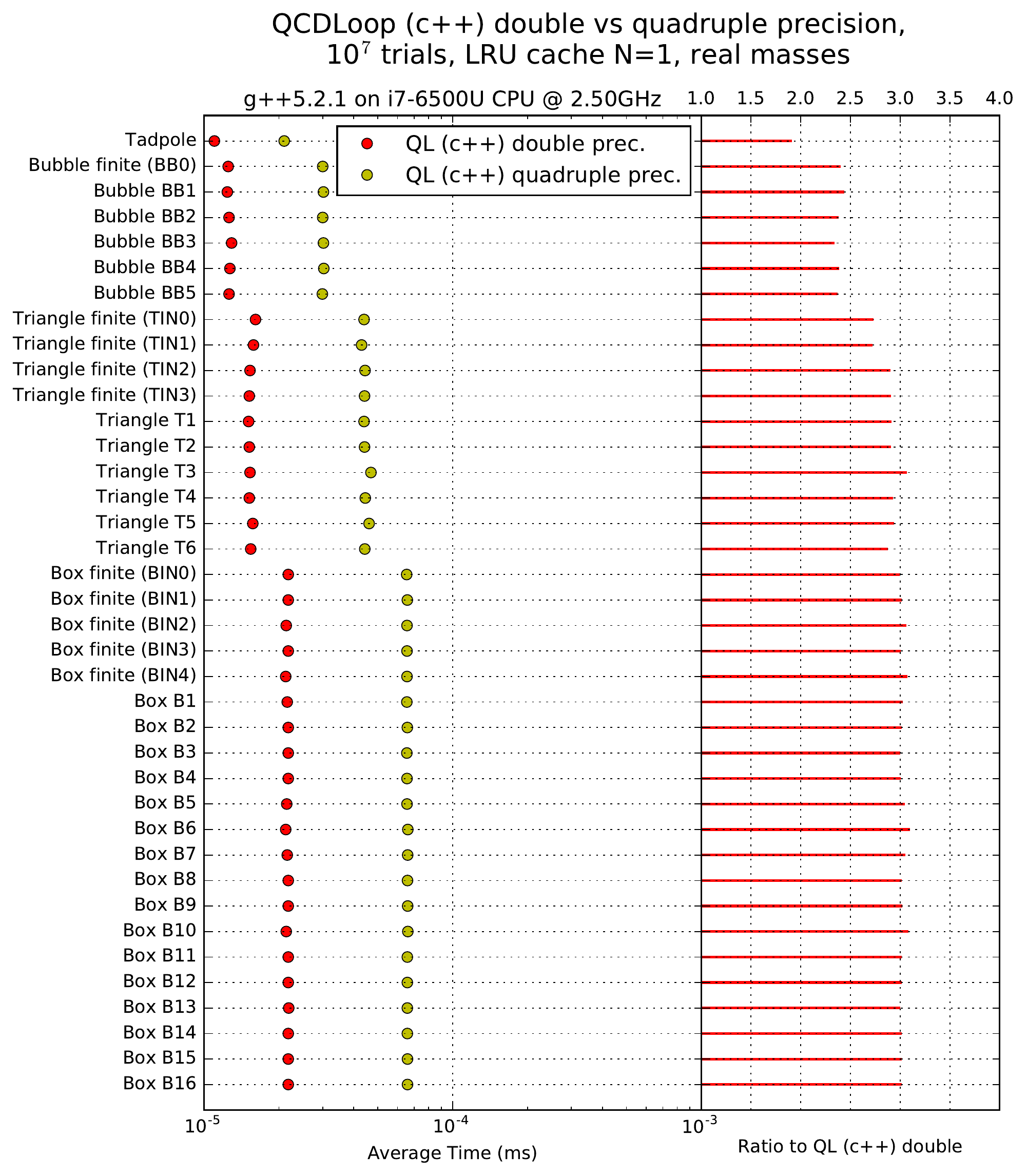}
\caption{\small Performance comparisons. Upper left: {\tt QCDLoop 2.0}
  vs {\tt QCDLoop 1.96} vs {\tt OneLoop 3.6} performance for real
  masses and double precision for $10^7$ configurations. Upper right:
  {\tt QCDLoop 2.0} vs {\tt QCDLoop 1.96} cache $N=1$ with real masses
  and double precision, for $10^7$ trials of the same
  configuration. Lower left: {\tt QCDLoop 2.0} cache $N>1$ vs no cache
  for $10^7$ trials of ten configurations. Lower right: {\tt QCDLoop
    2.0} double vs quadruple precision performance for $10^7$ trials of
  the same configuration.}
\label{fig:comparison}
\end{figure}
%%%%%%%%%%%%%%%%%%%%%%%%%%%%%%%%

In Figure~\ref{fig:comparison} we present four performance tests. In
the upper left plot we compare the two versions of {\tt QCDLoop}, {\it
  i.e.} the new {\tt c++} library {\tt QCDLoop 2.0} (red triangles)
and the previous {\tt fortran} library {\tt QCDLoop 1.96} (yellow
circles), to {\tt OneLoop 3.6} (blue boxes). Computations are
performed with disabled cache over $10^7$ random configurations using
real masses and double precision accuracy. Results are quoted in terms
of average time in milliseconds for each topology and specific
kinematics, following the notation of Table~\ref{tab:topo}. The right
inset shows the ratio to {\tt QCDLoop 1.96} where we notice that, on
average, the new library provides the best performance for tadpole,
bubble and triangle integrals, but we obtain similar timings to {\tt
  OneLoop} for box configurations. Overall we conclude that the new
library provides an improvement in comparison to past releases, in
particular when considering finite boxes computed in {\tt QCDLoop
  1.96} through the {\tt ff} library.  The upper right plot compares
the LRU cache between the two versions of {\tt QCDLoop}. We perform
$10^7$ trials of the same configuration for all topologies using real
masses and double precision. The new library cache is 20 to 30\%
faster than the previous one for tadpoles, bubbles and triangles,
meanwhile we observe a consistent speed-up of 70\% for boxes.  In the
lower left plot we compare the performance of the new {\tt QCDLoop}
library when using the LRU cache with $N=10$ in comparison to the
direct computation of 10 configurations repeated $10^7$ times. Results
show a great performance improvement when using this caching
mechanism. For some simple topologies like the tadpole the caching
algorithm has similar performance in comparison to the direct
computation, however when considering the most time consuming
configurations, like BIN4, differences of a factor 40 are observed.
Finally, in the lower right plot we compare the performance of the new
{\tt QCDLoop} library when computing results in double and quadruple
precision. In this case the same configuration is repeated $10^7$
times with double and quadruple variables. Differences are
proportional to the complexity of the integral. We always observe a
slowdown factor in the range 1.8 to 3 when using quadruple precision.

\begin{figure}[t]
\centering
\includegraphics[scale=0.4]{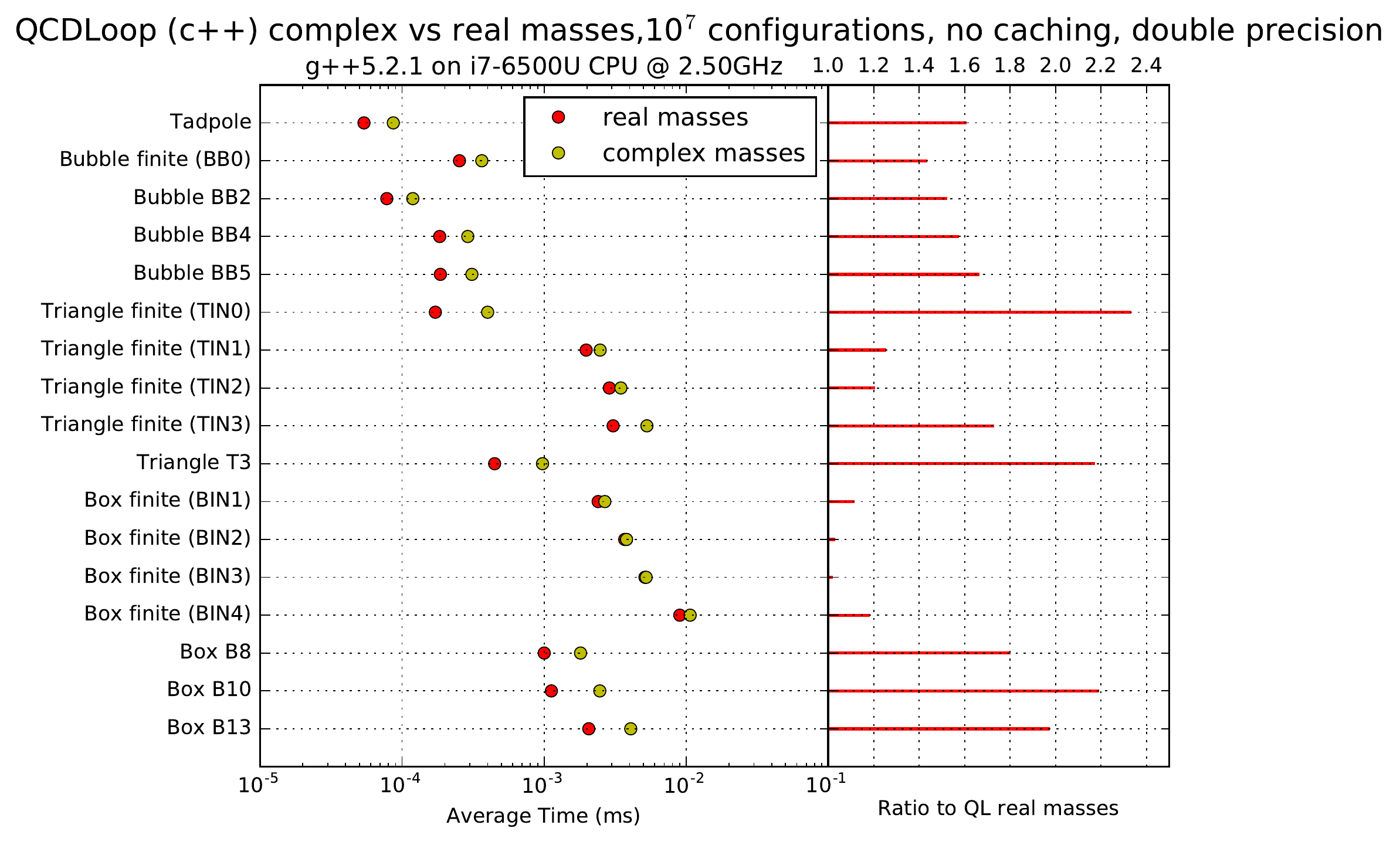}
\caption{\small {\tt QCDLoop} real vs complex masses performance.}
\label{fig:comparison2}
\end{figure}

In Figure~\ref{fig:comparison2} we compare the performance of the new
{\tt QCDLoop} library when computing one-loop scalar integrals with
real and complex masses. We observe a performance loss above 50\% when
activating complex masses for non finite integrals. Differences are
smaller for finite triangles and boxes due to the fact that those
implementations rely on complex objects for both mass types.

%%%%%%%%%%%%%%%%%%%%%%%%%%%%%%%%%%%%%%%%%%%%%%%%%%%%%%%%%%
%%%%%%%%%%%%%%%%%%%%%%%%%%%%%%%%

\subsection{Phenomenological applications}

In this section we test the new {\tt QCDLoop} library in Monte Carlo
environments. We first show results for real masses using simulations
obtained with the {\tt MCFM
  7.0.1}~\cite{Campbell:2000bg,Campbell:2012ft} interface. We then
test complex masses using interfaces to {\tt
  Ninja}~\cite{Mastrolia:2012bu,Peraro:2014cba} and {\tt
  GoSam}~\cite{vanDeurzen:2013saa} in a {\tt Sherpa
  2.2.0}~\cite{Hoche:2015sya,Hoche:2014kca} simulation.

\subsubsection{{\tt MCFM} interface}

The interface to the new {\tt QCDLoop} library and {\tt MCFM} is
straightforward, thanks to the backward compatible {\tt fortran}
wrapper presented in Section~\ref{sec:wrappers}.  The only technical
requirement in {\tt MCFM} consists in editing the {\tt makefile} and
updating the links and paths to the new library.

Simulations are performed for the LHC setup at $\sqrt{s}=13$ TeV,
using NNPDF3.0 NLO~\cite{Ball:2014uwa} set of PDFs and the default
parameters of the {\tt MCFM 7.0.1} input card. Here we focus on the
predictions of three processes which cover a large range of
topologies, including massive and massless loops, namely: $WW$ ({\tt
  nproc=61}), $ZZ$ ({\tt nproc=81}) and $\gamma\gamma\gamma\gamma$
({\tt nproc=289}) production. The aim of the results presented here is
to show that fully compatible results are obtained when running the MC
simulation with both versions of the library, {\it i.e.} {\tt QCDLoop
  1.96} and {\tt 2.0}.

\begin{figure}[t]
\centering
\includegraphics[scale=0.6]{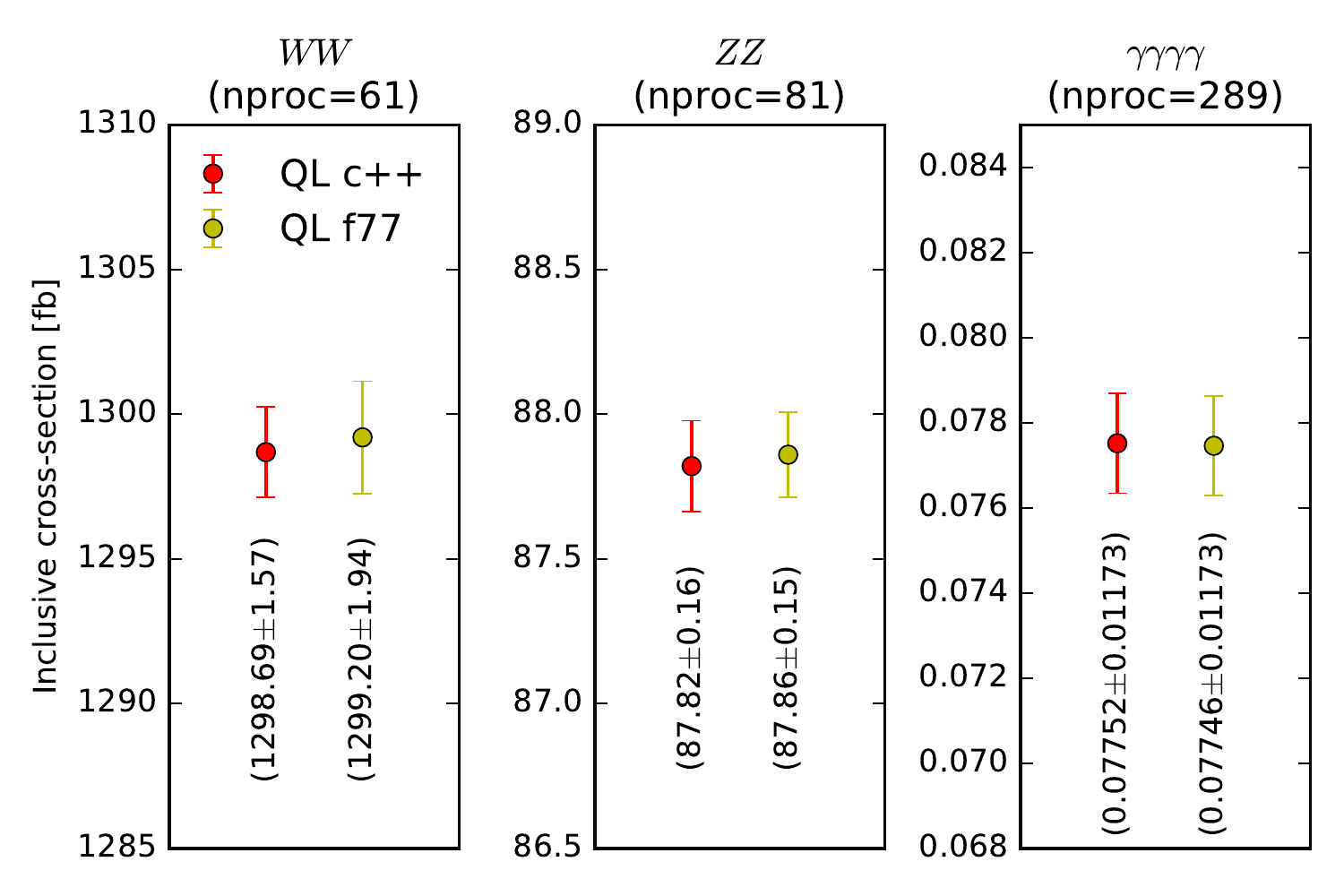}
\includegraphics[scale=0.6]{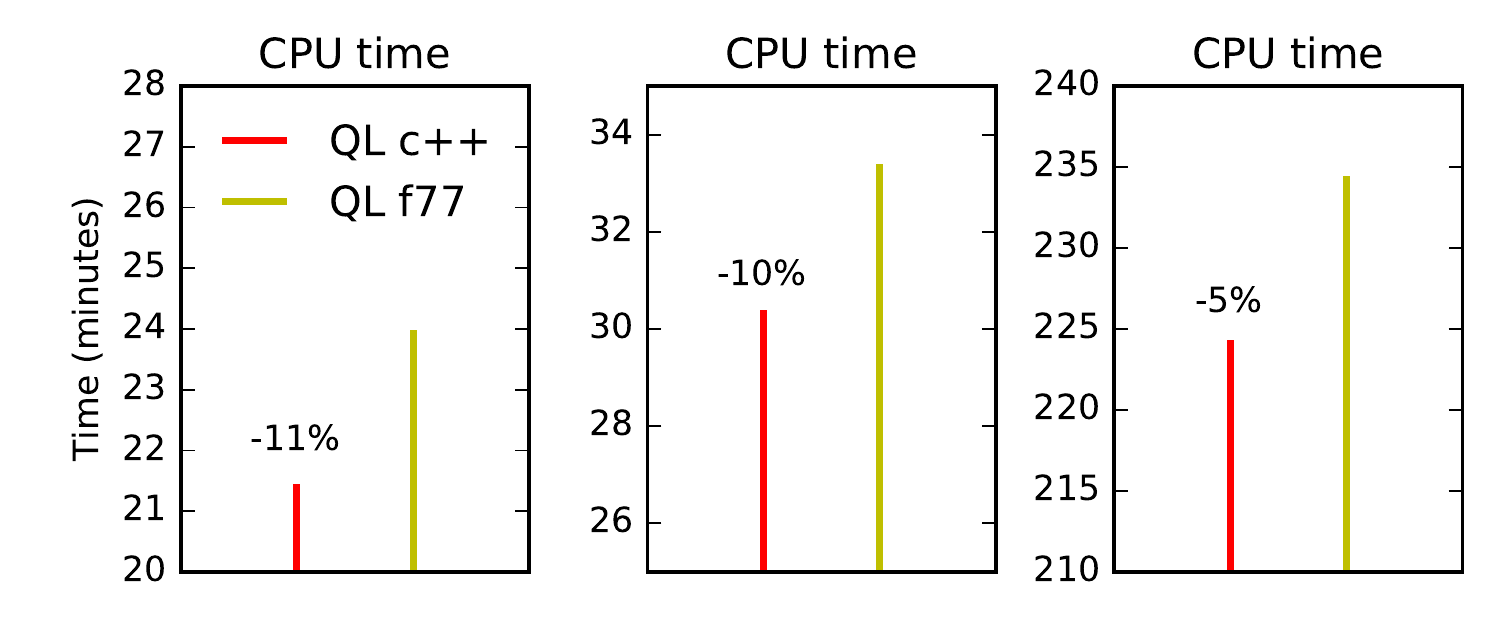}
\caption{\small Examples of inclusive cross-sections obtained with
  {\tt MCFM 7.0.1} using both versions of {\tt QCDLoop} for $WW$ ({\tt
    nproc=61}), $ZZ$ ({\tt nproc=81}) and $\gamma\gamma\gamma\gamma$
  ({\tt nproc=289}) production. Simulations performed for LHC @ 13
  TeV, using NNPDF3.0 NLO.}
\label{fig:mcfm}
\end{figure}

In Figure~\ref{fig:mcfm} we plot the inclusive cross-section for $WW$,
$ZZ$ and $\gamma\gamma\gamma\gamma$ processes respectively for both
versions of {\tt QCDLoop}. Numerical results are in agreement for all
processes. In terms of performance we observe 10\% improvement with
the new library for $WW$ and $ZZ$ production meanwhile 5\% improvement
for $\gamma\gamma\gamma\gamma$ production.

\begin{figure}[t]
\centering
\includegraphics[scale=0.4]{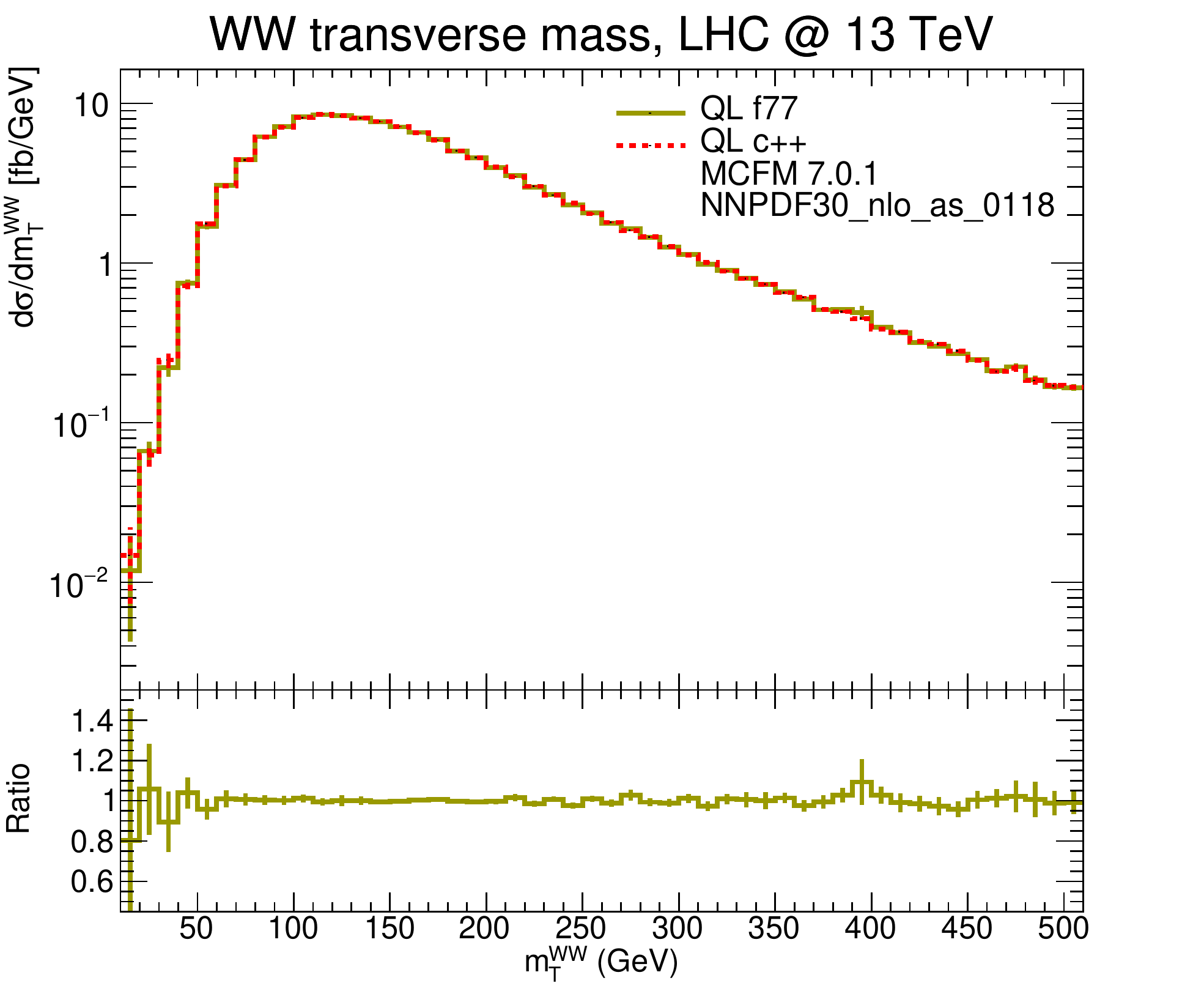}\includegraphics[scale=0.4]{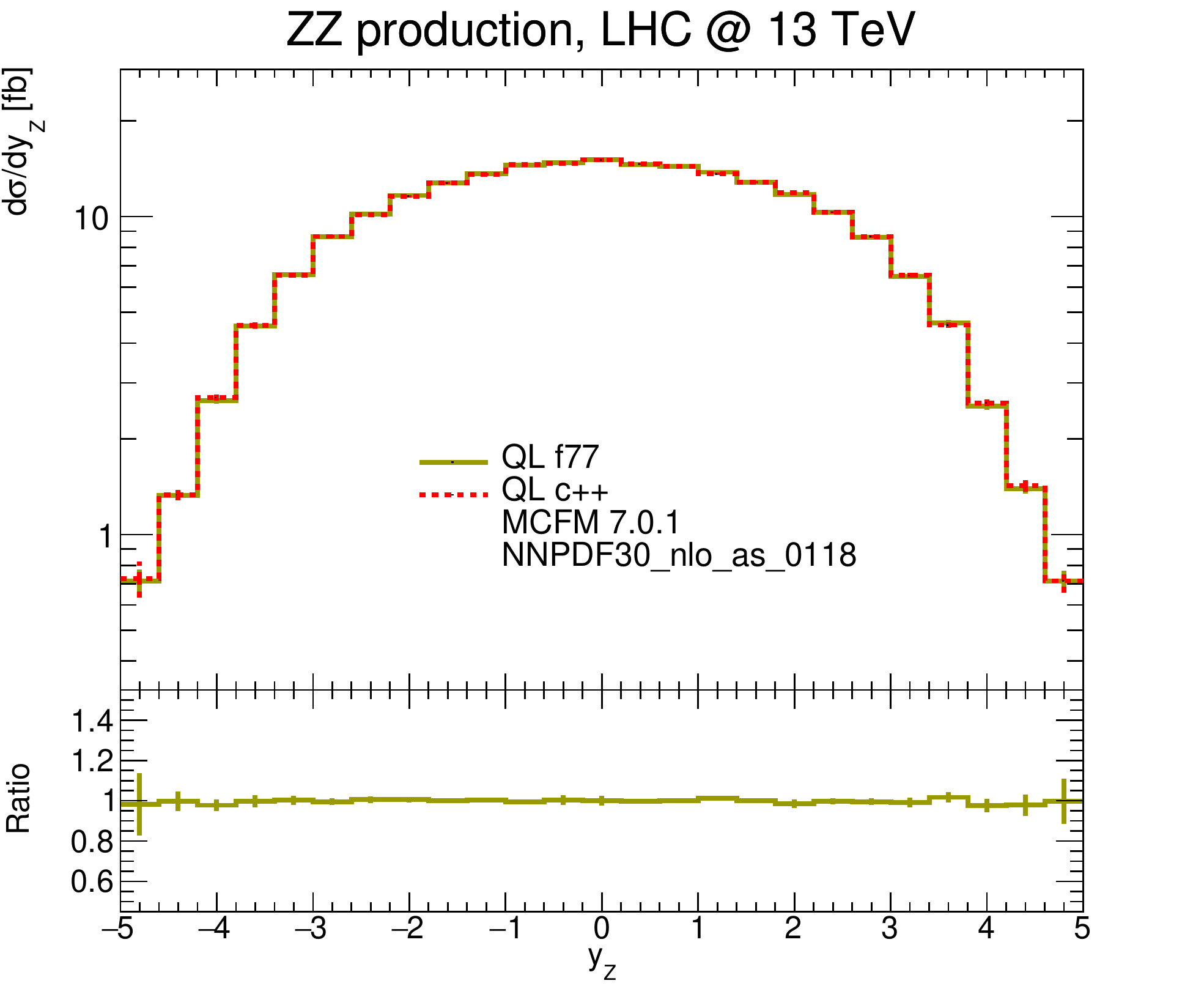}
\includegraphics[scale=0.4]{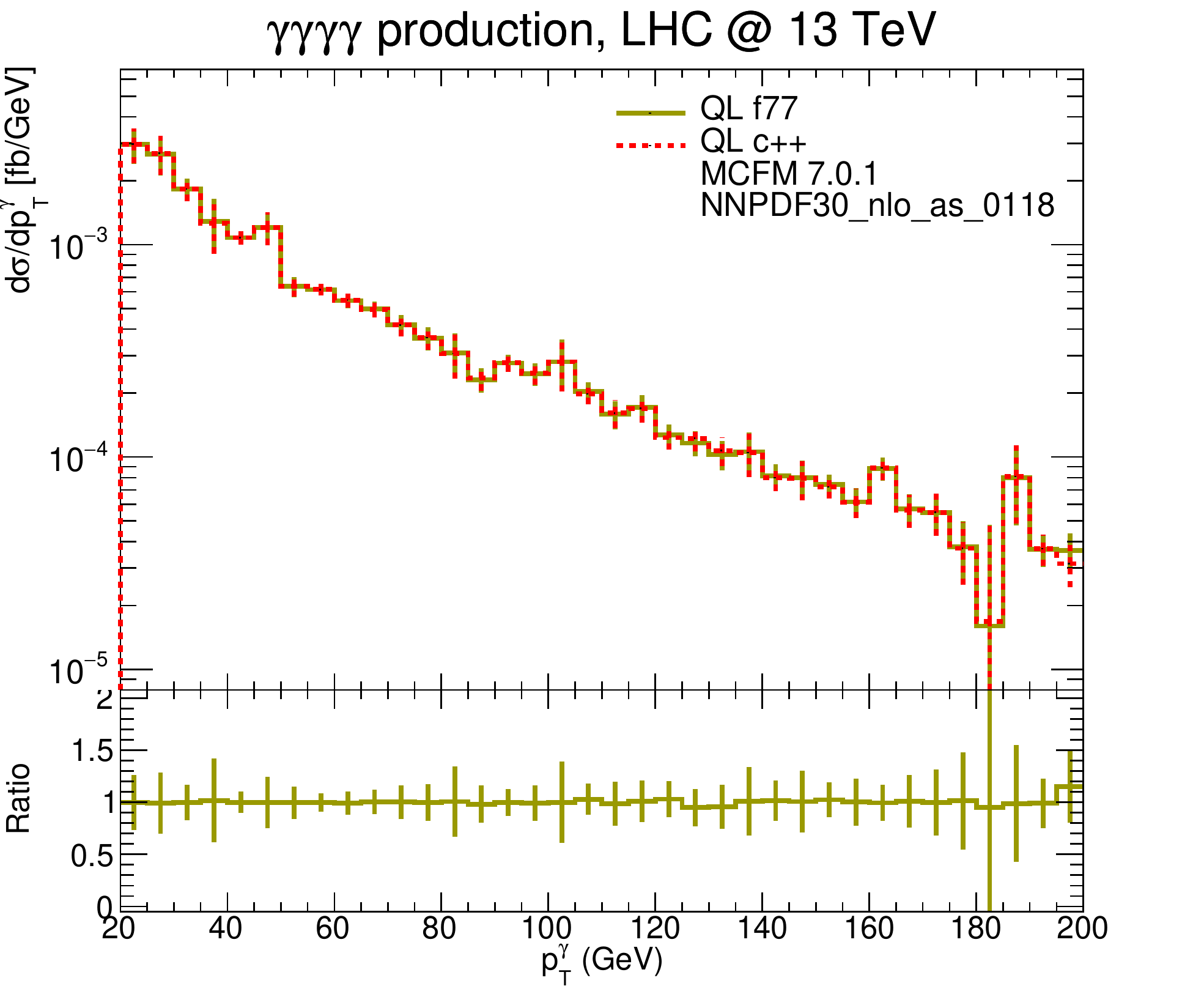}
\caption{\small Examples of differential distributions obtained with
  {\tt MCFM 7.0.1} using both versions of {\tt QCDLoop}, for the
  $m_{T}^{WW}$ transverse mass in $WW$ production, the average $y_Z$
  rapidity distribution in $ZZ$ production and the average
  $p^\gamma_{T}$ distribution in 4-$\gamma$ production. Simulations
  performed for LHC @ 13 TeV, using NNPDF3.0 NLO.}
\label{fig:mcfm2}
\end{figure}

In Figure~\ref{fig:mcfm2} we show differential distributions for the
three processes described above, always comparing both versions of the
{\tt QCDLoop} library. In the top left panel we show the $WW$-pair
transverse mass, $m_{T}^{WW}$, distribution. The top right panel presents
the rapidity of the lepton-pair $y_Z$ for the $ZZ$
production. Finally, the bottom plot highlights the photon
$p^{\gamma}_{T}$ distribution in the $\gamma\gamma\gamma\gamma$
production. In all cases the agreement is very good; for the
$p^{\gamma}_{T}$ distribution we have performed a simulation of few
hours (low statistics) in order to check that even with a low number
of iterations the final agreement between both codes is excellent.

%%%%%%%%%%%%%%%%%%%%%%%%%%%%%%%%%%%%%%%%%%%%%%%%%%%%%%%%%%
%%%%%%%%%%%%%%%%%%%%%%%%%%%%%%%%

%%%%%%%%%%%%%%%%%%%%%%%%%%%%%%%%%%%%%%%%%%%%%%%%%%%%%%%%%%
%%%%%%%%%%%%%%%%%%%%%%%%%%%%%%%%

\subsubsection{{\tt Ninja} and {\tt GoSam} interface}

For the validation of complex masses in a Monte Carlo environment we
considered the {\tt Ninja}
library~\cite{Mastrolia:2012bu,Peraro:2014cba}, which provides the
integrand reduction via Laurent expansion method for the computation
of one-loop integrals, and the {\tt GoSam}~\cite{vanDeurzen:2013saa}
automated package.

First, we expanded the {\tt Ninja} library with the new {\tt QCDLoop}
interface. We then verified the consistency of the new interface by
comparing the output between {\tt OneLoop 3.6} and {\tt QCDLoop 2.0}
for the examples provided by the {\tt Ninja} test codes. Second,
this new version of {\tt Ninja} was linked to {\tt GoSam} granting
access of matrix elements to Monte Carlo simulation tools.

\begin{figure}[t]
\centering
\includegraphics[scale=0.4]{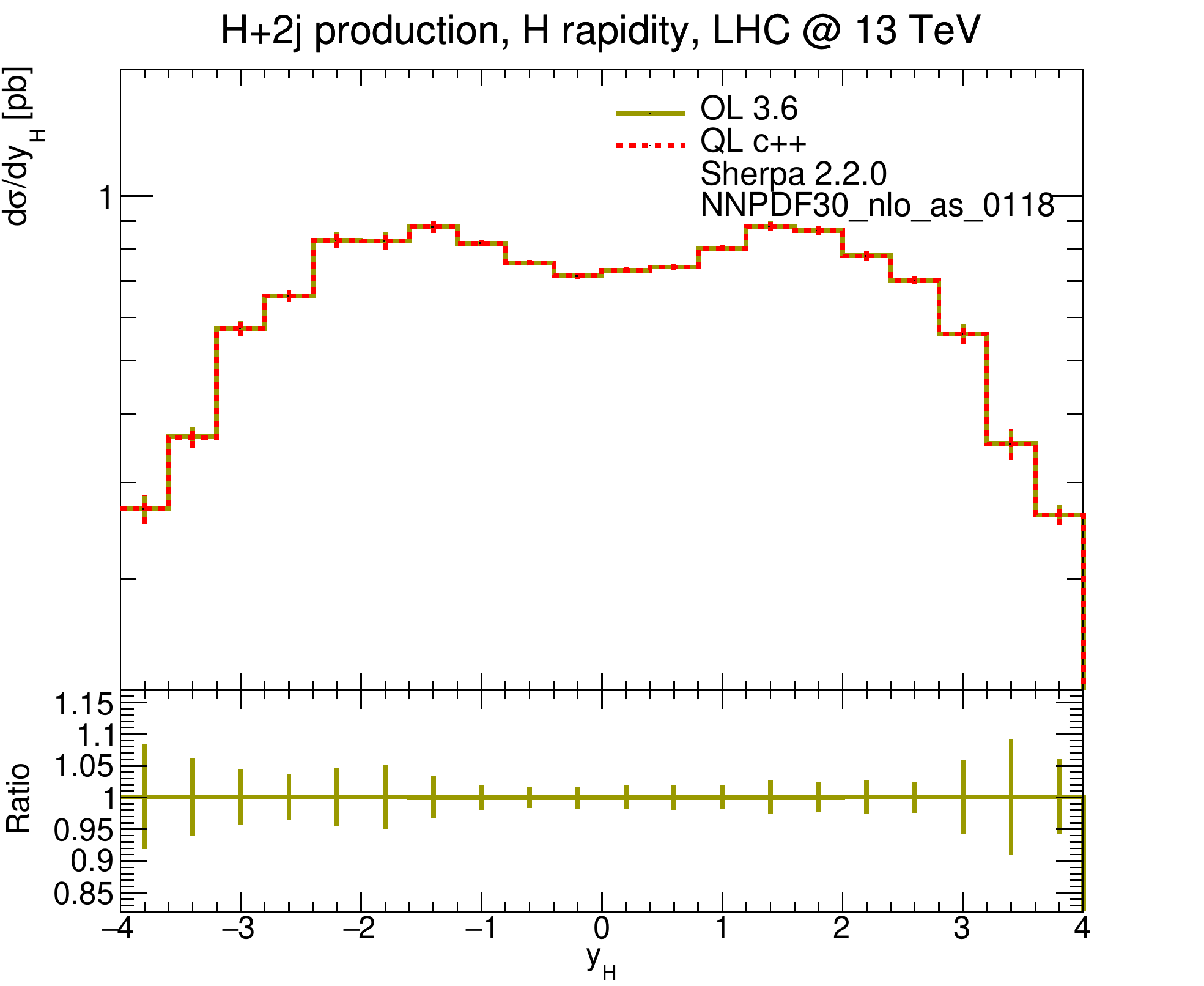}\includegraphics[scale=0.4]{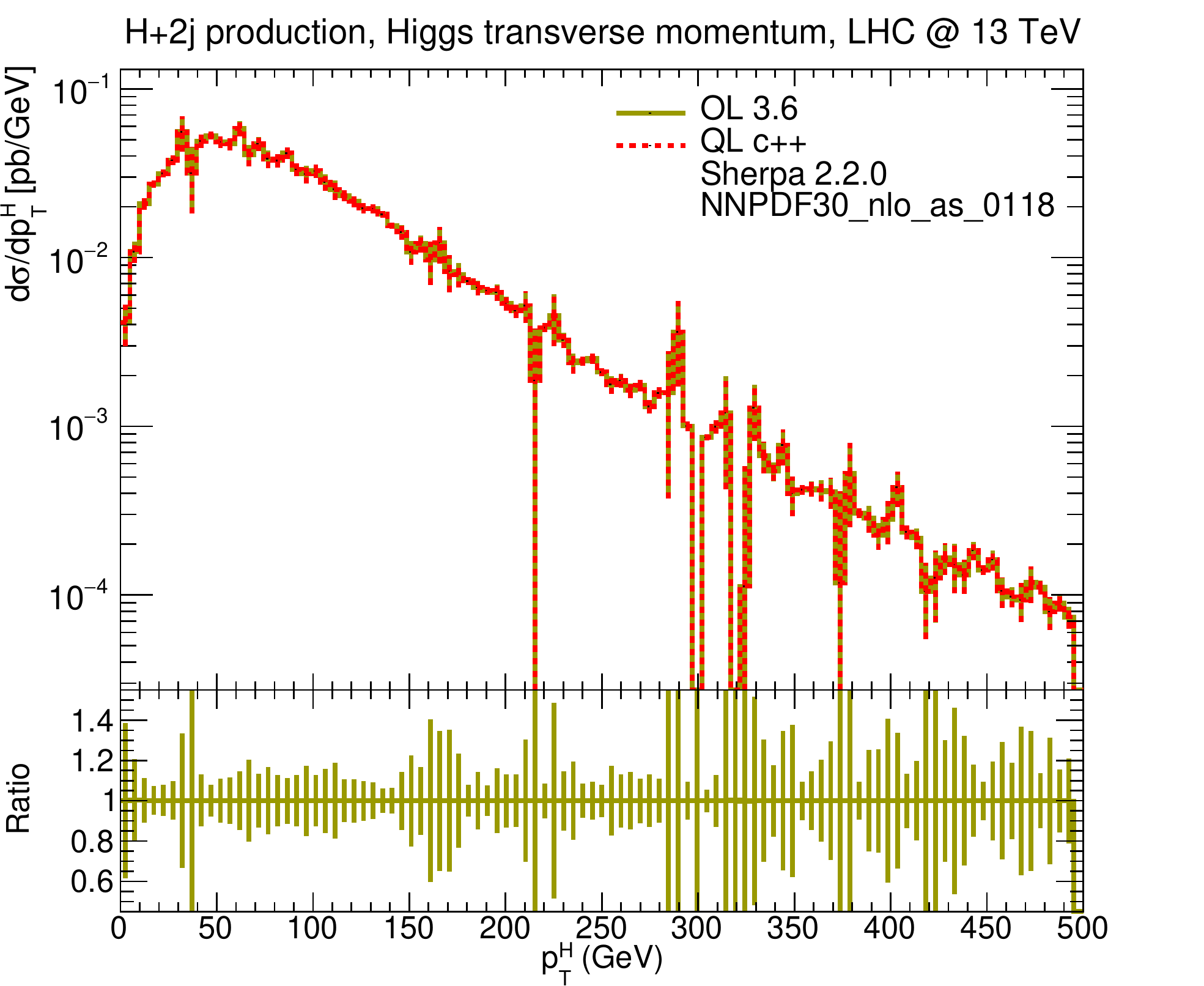}
\caption{\small Examples of differential distribution obtained with
  {\tt Sherpa 2.2.0} using {\tt Oneloop} and {\tt QCDLoop} through the
  {\tt Ninja/GoSam} interfaces. Simulations performed for LHC @ 13
  TeV, using NNPDF3.0 NLO.}
\label{fig:h2j}
\end{figure}

In order to provide quantitative results we used {\tt Sherpa 2.2.0} to
simulate $H+2j$ process at NLO, for the LHC setup at $\sqrt{s}=13$ TeV
with the NNPDF3.0 NLO PDF set. Such process is interesting for our
tests because it calls several topologies with complex masses. We
performed two simulations, the first with {\tt OneLoop} and the second
with {\tt QCDLoop}. Table~\ref{tab:xsh2j} shows the inclusive
cross-section values obtained with both codes with 50M events, and
Figure~\ref{fig:h2j} presents the corresponding rapidity $y_H$ and
$p_T^H$ distributions for the Higgs boson. This simulation shows that
numerical results are in agreement for all distributions within Monte
Carlo uncertainties. In terms of performance, both codes require $\sim
10$ CPU hours to complete the simulation.

\begin{table}[h]
  \begin{centering}
    \begin{tabular}{|c|c|}
      \hline 
      Library & $H+2j$ cross-section ({\tt Sherpa 2.2.0})\tabularnewline
      \hline 
      \hline 
      {\tt OneLoop 3.6} & $5.5867\pm(0.0121=0.21\%)$ pb\tabularnewline
      \hline 
      {\tt QCDLoop 2.0} & $5.5838\pm(0.0121=0.21\%)$ pb\tabularnewline
      \hline 
    \end{tabular}
    \par\end{centering}    
    \caption{\small Inclusive cross-section for $H+2j$ at NLO obtained
      with 5M events from {\tt Sherpa 2.2.0} interfaced with {\tt
        OneLoop 3.6} and {\tt QCDLoop 2.0} through the {\tt Ninja} and
      {\tt GoSam} interfaces.}
    \label{tab:xsh2j}
\end{table}

Finally, other interfaces to Monte Carlo codes are possible, users are
invited to interface their own code with {\tt QCDLoop}.

%%%%%%%%%%%%%%%%%%%%%%%%%%%%%%%%%%%%%%%%%%%%%%%%%%%%%%%%%%
%%%%%%%%%%%%%%%%%%%%%%%%%%%%%%%%

\section{Conclusions}
\label{sec-conclusion}

In this work we presented a new object-oriented framework for the 
{\tt QCDLoop} library. The new features compared with the
fortran version are
\begin{itemize}
\item
{\tt QCDLoop 2.0} calculates all integrals using native implementations.
The reliance on the external library {\tt ff} is no longer present. 
\item 
Full implementation of double and quadruple precision
for the whole library, including the possibility of switching 
between the two dynamically. This can be useful in regions of phase
space in which double precision is not sufficient. This can occur 
in corners of phase space, for example, in the context of NNLO calculations
when unresolved regions of phase space are probed.
\item
Improvements in the evaluation time with respect to the Fortran version.
\item
Improvements in the caching algorithm. For certain applications, in which 
the same integrals are needed several times, this can lead to 
great improvements in evaluation time with respect to the Fortran version.
In the new version one can adjust the size of the cache to yield the best 
performance.
\end{itemize}

The new {\tt QCDLoop} library is publicly available from the webpage:
\begin{center}
\url{http://cern.ch/qcdloop}
\end{center}
where instructions on how to install and run the code are also provided.

\section*{Acknowledgments}

We thank Simone Alioli, Gavin Salam for interesting discussions about
programming techniques, Fabrizio Caola for complex mass literature,
Tiziano Peraro for discussions about the {\tt Ninja} interface,
Gionata Luisoni for the {\tt GoSam} interface and the {\tt Sherpa}
setup presented in Sect.~\ref{sec-benchmarks}, and Ciaran Williams and
Tobias Neumann for testing. S.~C. and G.~Z. are supported by the
HICCUP ERC Consolidator grant (614577). R.~K.~E. acknowledges support
from the Alexander von Humboldt Foundation.

\clearpage

\bibliography{qcdloop}

\begin{thebibliography}{10}

\bibitem{Ellis:2011cr}
R.K. Ellis et~al.,
\newblock Phys. Rept. 518 (2012) 141, 1105.4319.
%%CITATION = ARXIV:1105.4319;%%

\bibitem{Andersen:2014efa}
J.R. Andersen et~al.,
\newblock (2014), 1405.1067.
%%CITATION = ARXIV:1405.1067;%%

\bibitem{Berger:2008sj}
C.F. Berger et~al.,
\newblock Phys. Rev. D78 (2008) 036003, 0803.4180.
%%CITATION = ARXIV:0803.4180;%%

\bibitem{Cullen:2011ac}
G. Cullen et~al.,
\newblock Eur. Phys. J. C72 (2012) 1889, 1111.2034.
%%CITATION = ARXIV:1111.2034;%%

\bibitem{Cascioli:2011va}
F. Cascioli, P. Maierhofer and S. Pozzorini,
\newblock Phys. Rev. Lett. 108 (2012) 111601, 1111.5206.
%%CITATION = ARXIV:1111.5206;%%

\bibitem{Alwall:2014hca}
J. Alwall et~al.,
\newblock JHEP 07 (2014) 079, 1405.0301.
%%CITATION = ARXIV:1405.0301;%%

\bibitem{Ellis:2007qk}
R.K. Ellis and G. Zanderighi,
\newblock JHEP 02 (2008) 002, 0712.1851.
%%CITATION = ARXIV:0712.1851;%%

\bibitem{vanOldenborgh:1990yc}
G.J. van Oldenborgh,
\newblock Comput. Phys. Commun. 66 (1991) 1.
%%CITATION = CPHCB,66,1;%%

\bibitem{Hahn:2006qw}
T. Hahn and M. Rauch,
\newblock Nucl. Phys. Proc. Suppl. 157 (2006) 236, hep-ph/0601248.
%%CITATION = HEP-PH/0601248;%%

\bibitem{vanHameren:2010cp}
A. van Hameren,
\newblock Comput. Phys. Commun. 182 (2011) 2427, 1007.4716.
%%CITATION = ARXIV:1007.4716;%%

\bibitem{Denner:2016kdg}
A. Denner, S. Dittmaier and L. Hofer,
\newblock (2016), 1604.06792.
%%CITATION = ARXIV:1604.06792;%%

\bibitem{Denner:2006ic}
A. Denner and S. Dittmaier,
\newblock Nucl. Phys. Proc. Suppl. 160 (2006) 22, hep-ph/0605312.
%%CITATION = HEP-PH/0605312;%%

\bibitem{Denner:2005nn}
A. Denner and S. Dittmaier,
\newblock Nucl. Phys. B734 (2006) 62, hep-ph/0509141.
%%CITATION = HEP-PH/0509141;%%

\bibitem{Denner:1991kt}
A. Denner,
\newblock Fortsch. Phys. 41 (1993) 307, 0709.1075.
%%CITATION = ARXIV:0709.1075;%%

\bibitem{'tHooft:1978xw}
G. 't~Hooft and M.J.G. Veltman,
\newblock Nucl. Phys. B153 (1979) 365.
%%CITATION = NUPHA,B153,365;%%

\bibitem{vanOldenborgh:1989wn}
G.J. van Oldenborgh and J.A.M. Vermaseren,
\newblock Z. Phys. C46 (1990) 425.
%%CITATION = ZEPYA,C46,425;%%

\bibitem{Denner:1991qq}
A. Denner, U. Nierste and R. Scharf,
\newblock Nucl. Phys. B367 (1991) 637.
%%CITATION = NUPHA,B367,637;%%

\bibitem{Campbell:2000bg}
J.M. Campbell and R.K. Ellis,
\newblock Phys. Rev. D62 (2000) 114012, hep-ph/0006304.
%%CITATION = HEP-PH/0006304;%%

\bibitem{Campbell:2012ft}
J.M. Campbell, H.B. Hartanto and C. Williams,
\newblock JHEP 11 (2012) 162, 1208.0566.
%%CITATION = ARXIV:1208.0566;%%

\bibitem{Mastrolia:2012bu}
P. Mastrolia, E. Mirabella and T. Peraro,
\newblock JHEP 06 (2012) 095, 1203.0291,
\newblock [Erratum: JHEP11,128(2012)].
%%CITATION = ARXIV:1203.0291;%%

\bibitem{Peraro:2014cba}
T. Peraro,
\newblock Comput. Phys. Commun. 185 (2014) 2771, 1403.1229.
%%CITATION = ARXIV:1403.1229;%%

\bibitem{vanDeurzen:2013saa}
H. van Deurzen et~al.,
\newblock JHEP 03 (2014) 115, 1312.6678.
%%CITATION = ARXIV:1312.6678;%%

\bibitem{Hoche:2015sya}
S. H{\"o}che and S. Prestel,
\newblock Eur. Phys. J. C75 (2015) 461, 1506.05057.
%%CITATION = ARXIV:1506.05057;%%

\bibitem{Hoche:2014kca}
S. H{\"o}che et~al.,
\newblock Eur. Phys. J. C75 (2015) 135, 1412.6478.
%%CITATION = ARXIV:1412.6478;%%

\bibitem{Ball:2014uwa}
NNPDF, R.D. Ball et~al.,
\newblock JHEP 04 (2015) 040, 1410.8849.
%%CITATION = ARXIV:1410.8849;%%

\end{thebibliography}

\end{document}